\documentclass[journal]{IEEEtran}

\usepackage{amssymb}
\usepackage{amsmath}
\usepackage{cite}
\usepackage{graphicx}
\usepackage{booktabs}
\usepackage{multirow}
\usepackage[normalem]{ulem}
\usepackage{tabularray}
\usepackage[ruled,linesnumbered]{algorithm2e}
\usepackage{xcolor}
\usepackage{bm}
\usepackage{hyperref}
\begin{document}

\title{DestripeCycleGAN: Stripe Simulation CycleGAN for Unsupervised Infrared Image Destriping}

\author{Shiqi~Yang,  
        Hanlin~Qin~\IEEEmembership{Member,~IEEE},
        Shuai~Yuan~\IEEEmembership{Student~Member,~IEEE},
        Xiang~Yan~\IEEEmembership{Member,~IEEE},
        and Hossein Rahmani

\thanks{This work was supported in part by the Shaanxi Province Key Research and Development Plan Project under Grant 2022JBGS2-09, in part by the Shaanxi Province Science and Technology Plan Project under Grant 2023KXJ-170, in part by the Xian City Science and Technology Plan Project under Grant 21JBGSZ-QCY9-0004, Grant 23ZDCYJSGG0011-2023, Grant 22JBGS-QCY4-0006, and Grant 23GBGS0001, in part by the Aeronautical Science Foundation of China under Grant 20230024081027, in part by the Natural Science Foundation Explore of Zhejiang province under Grant LTGG24F010001, in part by the Natural Science Foundation of Ningbo under Grant 2022J185, in part by the Technology Area Foundation of China 2021-JJ-1244, 2021-JJ-0471, 2023-JJ-0148, and in part by the China scholarship council 202306960052.
\textit{(Corresponding authors:~Hanlin Qin, Shuai Yuan.)}

Shiqi~Yang, Hanlin~Qin, Shuai~Yuan, and Xiang~Yan are with the School of Optoelectronic Engineering, Xidian University, Xi'an 710071, China (email: 
22191214967@stu.xidian.edu.cn; hlqin@mail.xidian.edu.cn; yuansy@stu.xidian.edu.cn; xyan@xidian.edu.cn).

Hossein Rahmani is with the School of Computing and Communications, Lancaster University, UK (email: h.rahmani@lancaster.ac.uk).
}
}

\markboth{Journal of \LaTeX\ Class Files,~VOL.~13, No.~9, September~2023}%
{Shell \MakeLowercase{\textit{et al.}}: Bare Demo of IEEEtran.cls for Journals}

\maketitle

\begin{abstract}
CycleGAN has been proven to be an advanced approach for unsupervised image restoration. This framework consists of two generators: a denoising one for inference and an auxiliary one for modeling noise to fulfill cycle-consistency constraints. 
However, when applied to the infrared destriping task, it becomes challenging for the vanilla auxiliary generator to consistently produce vertical noise under unsupervised constraints. 
This poses a threat to the effectiveness of the cycle-consistency loss, leading to stripe noise residual in the denoised image.
To address the above issue, we present a novel framework for single-frame infrared image destriping, named DestripeCycleGAN.
In this model, the conventional auxiliary generator is replaced with a priori stripe generation model (SGM) to introduce vertical stripe noise in the clean data, and the gradient map is employed to re-establish cycle-consistency.
Meanwhile, a Haar wavelet background guidance module (HBGM) has been designed to minimize the divergence of background details between the different domains.
To preserve vertical edges, a multi-level wavelet U-Net (MWUNet) is proposed as the denoising generator, which utilizes the Haar wavelet transform as the sampler to decline directional information loss. Moreover, it incorporates the group fusion block (GFB) into skip connections to fuse the multi-scale features and build the context of long-distance dependencies.
Extensive experiments on real and synthetic data demonstrate that our DestripeCycleGAN surpasses the state-of-the-art methods in terms of visual quality and quantitative evaluation. Our code will be made public at \url{https://github.com/0wuji/DestripeCycleGAN}.
\end{abstract}


\section{Introduction}
\IEEEPARstart{I}{nfrared} imaging systems have extensive applications, including autonomous driving~\cite{aizat2023comprehensive}, maritime rescue~\cite{wu2023mtu}, and reconnaissance activities~\cite{fang2022infrared}.
However, the different column response of the infrared focal plane array (IRFPA) always leads to irregular stripe noise in infrared images, which significantly affects the performance of downstream tasks such as target detection~\cite{yuan2024sctransnet} and image segmentation~\cite{Badrinarayanan2017Segnet}.
To mitigate this issue, numerous methods have been proposed. 
Traditional destriping methods mainly construct prior models between degraded image and background, including filters~\cite{cao2015effective, zeng2019fourier, zeng2020fourier}, data statistics~\cite{tendero2012non, horn1979destriping, wegener1990destriping} and optimization techniques~\cite{chang2016remote, song2023simultaneous, he2023fspnp}. 
Nevertheless, these methods always indiscriminately remove stripe noise and background details, leading to significant blurring of the vertical edges. 

Recently, data-driven methods have been widely proposed for learning stripe features in a supervised manner, which can distinguish the structure of stripe noise and background from a large receptive field. 
Early learning-based methods removed stripe noise by designing shallow convolutional neural networks (CNNs)~\cite{kuang2017single, xiao2018removing}. However, these methods struggle to efficiently capture the semantic information of both the stripe component and the image background, resulting in residual noise. 
To address the aforementioned issue, researchers try to construct deeper networks that leverage techniques such as residual learning to enhance the efficiency of feature representation~\cite{he2018single, li2021Non, jong2020dual}. 
Based on this, multi-scale fusion~\cite{chang2019infrared, xu2022single} and attention mechanisms~\cite{guan2020fixed, ding2022single, li2023progressive, yuan2024arcnet} have been proposed to differentiate stripes from the background globally. Most recently, a priori information about stripes, including orientation features and noise response properties, has been widely used to highlight the stripe component and thus preserve the background structure~\cite{wang2022noise, cao2022robust}. 
Among these methods, the wavelet transform has proven to be an effective tool for modeling stripe structure~\cite{munch2009stripe}, achieving multiscale representation~\cite{zhong2020partial}, and enabling lossless decomposition and recovery~\cite{chui1992introduction}. 
Therefore, many works combine the wavelet transform with CNNs to achieve superior performance~\cite{guan2019wavelet, chang2019toward, zhang2021research}. Unfortunately, the supervised learning paradigm needs pairs of background-matched stripy images and target images, which are difficult to obtain in real scenarios.


\begin{figure*}[t]
    \centering
    \includegraphics[width=1\textwidth]{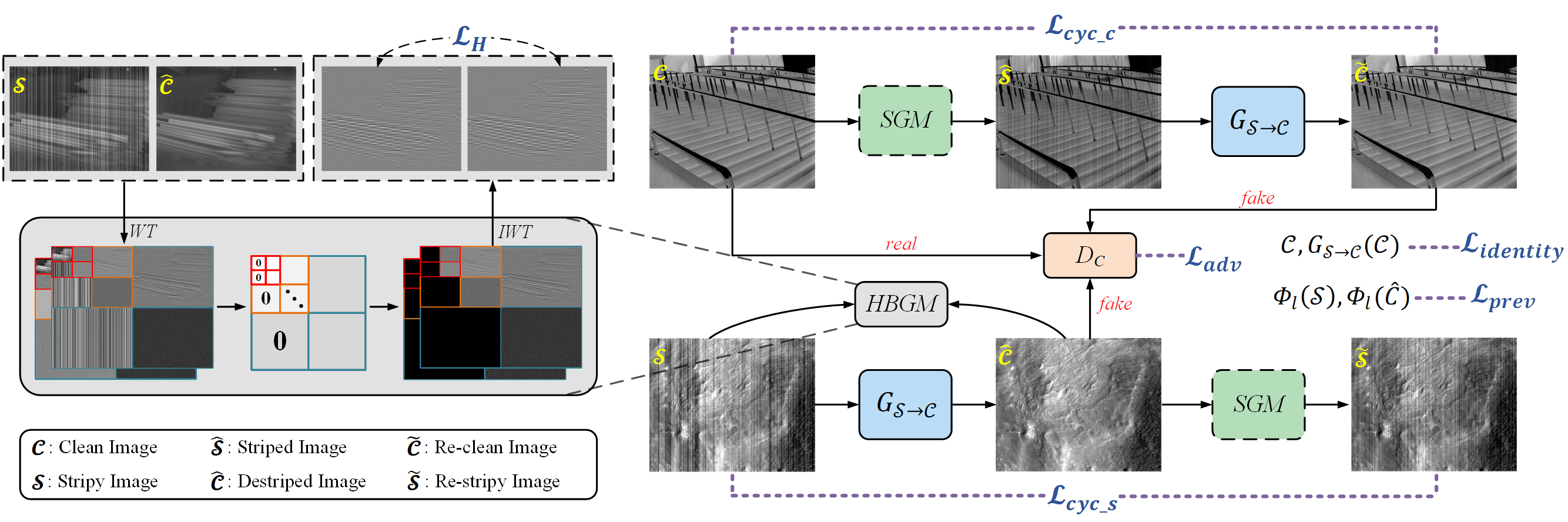}
    \caption{Overview of the proposed DestripeCycleGAN for infrared image destriping.
    Our DestripeCycleGAN mainly consists of three modules: stripe generation model (SGM), generator ($G_{\mathcal{S} \rightarrow \mathcal{C}}$), and Haar wavelet background guidance module (HBGM).
    SGM and $G_{\mathcal{S} \rightarrow \mathcal{C}}$ are employed to model the stripe noise and restore the clean image separately, jointly constructing the unsupervised architecture for $stripy \leftrightarrow clean$.
    HBGM is utilized to maintain background consistency between the stripy image and the destriped result. 
    Moreover, the discriminator $D_{\mathcal{C}}$ is used to build adversarial loss with $G_{\mathcal{S} \rightarrow \mathcal{C}}$.
    }
    \label{fig:1}
\end{figure*}

To effectively remove real noise, unsupervised learning has attracted much attention~\cite{isola2017image, zhu2017unpaired, lehtinen2018noise2noise, mei2018unsupervised}.
Among them, the most typical structure is CycleGAN~\cite{zhu2017unpaired}, which contains two generators: a denoising generator to recover the clean image from the noise domain and an auxiliary generator to fit real noise from the clean domain.
Owing to the powerful domain transformation capability of CycleGAN, it has been widely used in low-level vision tasks, including deraining, deflicker, and decluttering~\cite{song2020unsupervised, wei2021deraincyclegan, lin2023deflickercyclegan, liu2023toward}. These successful applications inspire us to naturally apply this framework to the infrared image destriping task.
However, observing that stripe noise exhibits prominent directional structural properties, applying the vanilla CycleGAN directly to the destriping task will result in disastrous outcomes.
This is because auxiliary generators, despite having strong fitting abilities, struggle to consistently produce the vertical stripe noise through unconstrained data-driven strategies.

To address this issue, we propose a DestripeCycleGAN, which replaces the auxiliary generator with a customized stripe generation model (SGM). Unlike the conventional auxiliary generator, our SGM can explicitly generate Gaussian-distributed vertical stripe noise with varying intensities during the training stage, guaranteeing the preservation of the directional characteristics of the stripe structure.
Correspondingly, considering there exist obvious distinctions in the horizontal distribution between the stripes generated by SGM and those in the original dirty map, we apply the cycle-consistency constraints to the vertical gradient map of the stripe image, rather than the original image.
Our framework possesses two advantages: (a) The incorporation of a structural prior of stripe noise in the SGM enables more accurate modeling of the semantic disparities between the target and noise domains.
(b) By replacing the redundant auxiliary generator, the SGM reduces the model parameters and computational complexity of DestripeCycleGAN compared to the vanilla CycleGAN.

As DestripeCycleGAN comprises only one learnable generator, utilizing a simple U-Net model is insufficient for precisely decoupling the stripe noise from complex backgrounds~\cite{song2020unsupervised, wei2021deraincyclegan, lin2023deflickercyclegan, liu2023toward}.
Consequently, we optimize the generator in three ways:
First, we design a Haar background guidance module (HBGM) astride the generator to facilitate the consistency of background details.
Following that, we propose a multi-level wavelet U-Net (MWUNet) as the generator, utilizing Haar wavelet transform to sample features to reduce the loss of intermediate features. 
Lastly, we incorporate a group fusion block (GFB) to merge multi-scale context and build long-range dependencies within skip connections of MWUNet.
The main contributions of this paper are as follows: 
\begin{itemize} 
    \item An efficient deep unsupervised DestripeCycleGAN is proposed to balance the semantic information between degraded and clean domains for infrared image destriping.

    \item We design a Haar wavelet background guidance module (HBGM) to eliminate the influence of vertical stripes, thereby ensuring the consistency of background details between the stripe and clean domain.
   
    \item We incorporate group fusion blocks (GFB) into the Multi-level Wavelet U-Net (MWUNet) as a novel generator, which can prevent feature loss during sampling, fuse multi-scale features, and enhance long-range dependencies.

\end{itemize}

\section{Method}
As shown in Fig.~\ref{fig:1}, DestripeCycleGAN constructs an end-to-end mapping between stripe domain ($\mathcal{S}$) and clean domain ($\mathcal{C}$).
Specifically, our framework consists of three basic components: 
(a) the stripe generation model (SGM) is utilized to simulate stripe noise on the clean domain; 
(b) the generator $G_{\mathcal{S} \rightarrow \mathcal{C}}$, which rebuilds the clean background from the stripe noise distribution; 
(c) a discriminator $D_{\mathcal{C}}$ to determine whether a clean domain is real or developed by $G_{\mathcal{S} \rightarrow \mathcal{C}}$. 
Given stripy and clean images, the DestripeCycleGAN process flow comprises two branches: \romannumeral1)~$\mathcal{C} \rightarrow \hat{\mathcal{S}} \rightarrow \widetilde{\mathcal{C}}$, which simulates stripes on the clean domain with SGM and uses $G_{\mathcal{S} \rightarrow \mathcal{C}}$ to reconstruct them; \romannumeral2)~$\mathcal{S} \rightarrow \hat{\mathcal{C}} \rightarrow \widetilde{\mathcal{S}}$, which removes stripe noise and recover the stripe domain.

\subsection{Stripe generation model (SGM)}
In the infrared image destriping task, the simulated stripe noise with actual noise characteristics is the foundation for designing the algorithm and testing the model. Researchers have proposed simple offset~\cite{kuang2017single}, linear~\cite{narayanan2005scene, liu2015fixed}, and nonlinear~\cite{he2018single, xu2022single} models to constract stripe noises in previous work. 
To better approximate the actual distribution of stripe noise, we employ a polynomial-based nonlinear noise model to construct the SGM and simulation dataset so that the paired clean images $\mathbf{I}_{\mathbf{C}}$ and stripe images $\mathbf{I}_{\mathbf{S}}$ are defined as:
\begin{equation}
\label{eq1}
\mathbf{I}_{\mathbf{S}} = \mathbf{I}_{\mathbf{C}} + \mathit{A}_{m}^{n} \mathbf{I}_{\mathbf{C}} + \mathit{A}_{m-1}^{n-1} \mathbf{I}_{\mathbf{C}} + \cdots + \mathit{A}_{1}\mathbf{I}_{\mathbf{C}} + \mathit{A}_{0}
\end{equation}
in which $A_{m}=(a_{ij})_{m}$ is the column-fixed polynomial coefficients matrix and $n$ represents the degree of the polynomial. In this paper, we define $a_{i}\sim N(0,\sigma^2)$ and $a_{i}\sim U(-\mu,\mu)$ as Gaussian-distributed and Uniform-distributed noise, respectively. The $\sigma$ and $\mu$ denote the corresponding noise intensity, and the $m$ is set to 3~\cite{he2018single}.

\begin{algorithm}[t]
\caption{SGM algorithm}\label{algorithm}
\label{alg1}
\KwData{clean images $\mathbf{I}_{\mathbf{C}} \in \mathbb{R}^{B \times H \times W \times C}$ from $\mathcal{C}$ or $\hat{\mathcal{C}}$, $B$ represents the number of training batches, and $C$, $H$, and $W$ represent the image's channel, height, and width, respectively.}
\KwResult{stripe images $\mathbf{I}_{\mathbf{S}} \in \mathbb{R}^{B \times H \times W \times C}$.}
\For{$k=1,2,3,\cdots, B$}{
\For{m=0, 1, 2, 3}{
$\sigma_{m}$ = Random(0.02, 0.12)\; \textcolor{olive}{$/*~\sigma_{m} \in \mathbb{R}^{1} ~*/$} \\
$(a_{i,:})_{m}=Normal(0, \sigma_{m}, W)$\;  \textcolor{olive}{$/*~(a_{i,:})_{m} \in \mathbb{R}^{W}~*/$} \\
$A_{m}=Repeat((a_{i,:})_{m}, H)$\; \textcolor{olive}{$/*~A_{m} \in \mathbb{R}^{H \times W}~*/$} \\
$expand\_dims(A_{m})$\; \textcolor{olive}{$/*~A_{m} \in \mathbb{R}^{H \times W \times 1}~*/$}
}
$\mathbf{I}_{\mathbf{S}_{k}} = \mathbf{I}_{\mathbf{C}_{k}} + \mathit{A}_{3}^{3} \mathbf{I}_{\mathbf{C}_{k}} \ + \mathit{A}_{2}^{2} \mathbf{I}_{\mathbf{C}_{k}} + \mathit{A}_{1}\mathbf{I}_{\mathbf{C}_{k}} + \mathit{A}_{0}$\;
}
\end{algorithm}

Based on the stripe noise model described in Eq.~\ref{eq1}, our SGM is proposed to consistently generate stripe images according to ~\textbf{Algorithm~\ref{alg1}}. 
Specifically, we model the SGM by a third-order Gaussian noise distribution and set $\sigma$ to a random value between $[0.02,0.12]$ for each batch to increase the model's generalization. 
Consequently, in the branch of $\mathcal{C} \rightarrow \hat{\mathcal{S}} \rightarrow \widetilde{\mathcal{C}}$, the SGM generates a simulated stripe image, which is then recovered as a clean image by the generator.
We utilize MS-SSIM mixing $L_{1}$ loss to constrain the cycle-consistency as follows~\cite{zhao2016loss}:
\begin{equation}
\label{eq2}
\begin{split}
\mathcal{L}_{cyc\_c} &= \mathbb{E}_{c\sim p(c)}[\alpha (1-MS\text{-}SSIM(\mathcal{C},\widetilde{\mathcal{C}}))\\
&+(1-\alpha )G_{\sigma _{G}^{m} } \mathcal{L}_{1}(\mathcal{C},\widetilde{\mathcal{C}})] 
\end{split}
\end{equation}
\begin{equation}
\label{eq3}
MS\text{-}SSIM(x,y)=\prod_{m=1}^{M} (\frac{\mu_{x} \mu _{y} +c_{1} }{\mu _{x}^{2} + \mu _{x}^{2} + c_{1}})
(\frac{2\sigma _{xy} +c_{2} }{\sigma  _{x}^{2} + \sigma  _{x}^{2} + c_{2}})
\end{equation}
In Eq. \ref{eq2}, $G_{\sigma _{G}^{m}}$ represents the Gaussian filter represents the Gaussian filter with various standard deviations $\sigma _{G}^{m}$. In Eq. \ref{eq3}, when given the image $x$, $y$, $G_{\sigma _{G}^{m}}$ is used to compute their mean value $\mu _{x}$, $\mu _{y}$, standard deviation $\sigma _{x}$, $\sigma _{y}$ and covariance $\sigma _{xy}$. 
M represents the number of scales, which we downsample four times thus M is set to 5. $c_{1}$, $c_{2}$ are constants used to prevent division by zero.


\begin{figure*}[t]
    \centering
    \includegraphics[width=0.7\textwidth]{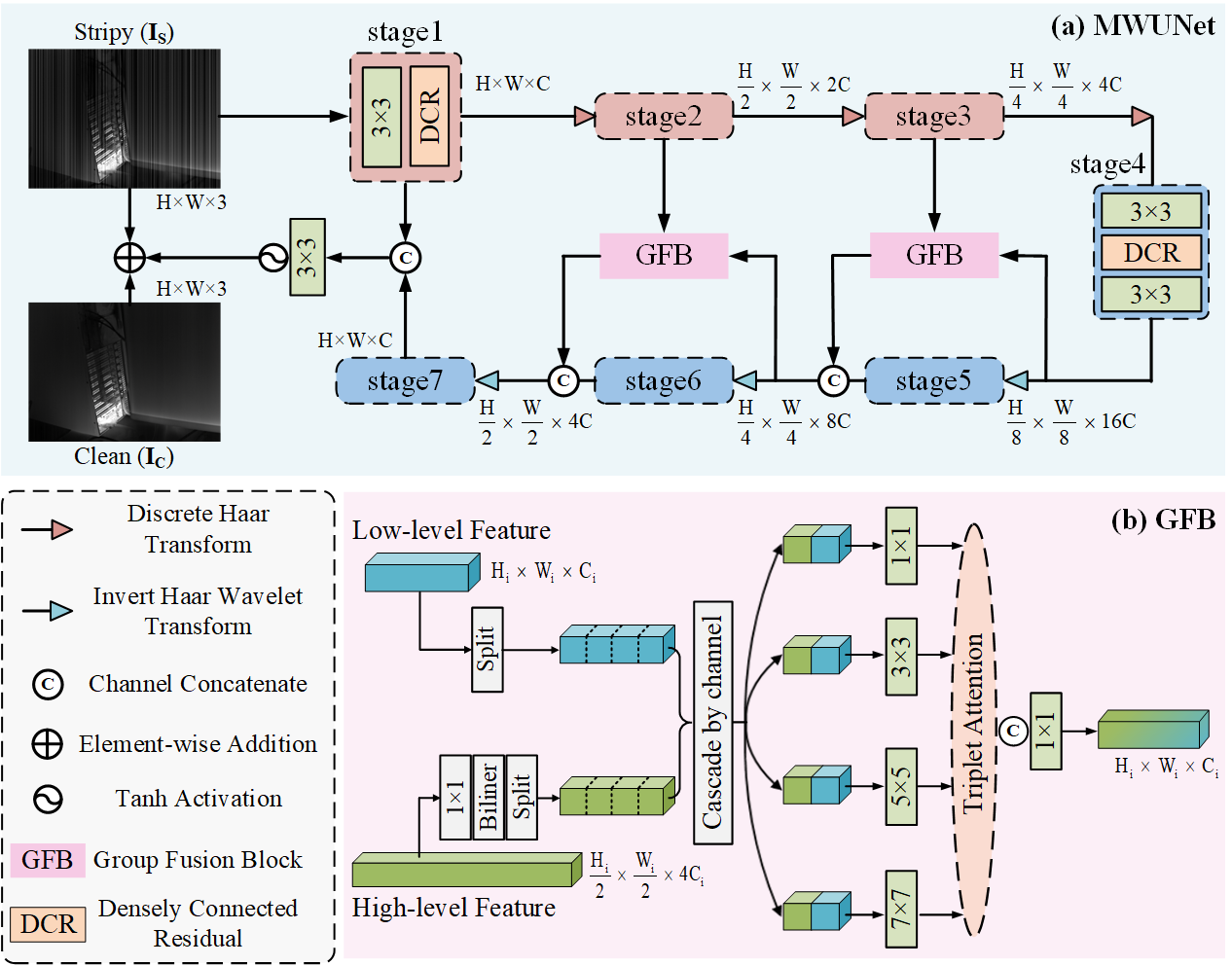}
    \caption{Detailed architecture of the generator $G_{\mathcal{S} \rightarrow \mathcal{C}}$: (a) Multi-level Wavelet U-Net (MWUNet) using the Harr wavelet transform as the sampler to reduce information loss, and (b) group fusion blocks (GFB) that incorporate multi-scale features into the skip connection.
    }
    \label{fig:3}
\end{figure*}

\begin{table*}[t]
\centering
\footnotesize
\caption{The PSNR/SSIM values of different methods for two types and two intensities of stripe noise across three test sets}
\label{table1}
\begin{tblr}{
  cells = {c},
  cell{2}{1} = {r=8}{},
  cell{2}{2} = {r=4}{},
  cell{2}{3} = {r=2}{},
  cell{4}{3} = {r=2}{},
  cell{6}{2} = {r=4}{},
  cell{6}{3} = {r=2}{},
  cell{8}{3} = {r=2}{},
  cell{10}{1} = {r=8}{},
  cell{10}{2} = {r=4}{},
  cell{10}{3} = {r=2}{},
  cell{12}{3} = {r=2}{},
  cell{14}{2} = {r=4}{},
  cell{14}{3} = {r=2}{},
  cell{16}{3} = {r=2}{},
  cell{18}{1} = {r=8}{},
  cell{18}{2} = {r=4}{},
  cell{18}{3} = {r=2}{},
  cell{20}{3} = {r=2}{},
  cell{22}{2} = {r=4}{},
  cell{22}{3} = {r=2}{},
  cell{24}{3} = {r=2}{},
  hline{1-2,10,18,26} = {-}{},
  hline{6,14,22} = {2-11}{},
}
\hline
Datasets  & Categorys & Intensity & GF                & LRSID    & SEID     & DLS-NUC  & DMRN     & SNRWDNN           & TSWEU             & Ours                     \\
DLS\_50   & Gaussian  & 0.05      & 34.4124/          & 31.7674/ & 35.2960/ & 30.6880/ & 30.5507/ & \textit{36.1078/} & \uline{37.0638/}  & \textbf{39.1159/}        \\
          &           &           & 0.9677            & 0.9470   & 0.9383   & 0.9415   & 0.9541   & \textit{0.9728}   & \uline{0.9813}    & \textbf{\textbf{0.9846}} \\
          &           & 0.1       & 30.6014/          & 28.8836/ & 30.1175/ & 26.8997/ & 27.0808/ & \textit{33.2277/} & \uline{35.6812/}  & \textbf{36.5678/}        \\
          &           &           & 0.9296            & 0.9357   & 0.8651   & 0.8605   & 0.9056   & \textit{0.9558}   & \uline{0.9745}    & \textbf{\textbf{0.9782}} \\
          & Uniform   & 0.05      & 38.5470/          & 32.4397/ & 36.5151/ & 32.9539/ & 29.8265/ & \uline{39.8972/}  & \textit{38.3174/} & \textbf{42.9974/}        \\
          &           &           & 0.9793            & 0.9507   & 0.9434   & 0.9521   & 0.9505   & \uline{0.9800}    & \textit{0.9646}   & \textbf{\textbf{0.9886}} \\
          &           & 0.1       & 34.2006/          & 30.2629/ & 34.1340/ & 30.6176/ & 29.6108/ & \textit{36.1353/} & \uline{37.7852/}  & \textbf{38.8657/}        \\
          &           &           & 0.9606            & 0.9440   & 0.9269   & 0.9391   & 0.9466   & \textit{0.9691}   & \uline{0.9831}    & \textbf{\textbf{0.9821}} \\
Set12     & Gaussian  & 0.05      & 32.0770/          & 28.4463/ & 34.2887/ & 28.1707/ & 29.2658/ & \textit{35.2049/} & \uline{35.2675/}  & \textbf{36.6098/}        \\
          &           &           & 0.9693            & 0.9623   & 0.9514   & 0.9404   & 0.9726   & \textit{0.9828}   & \uline{0.9830}    & \textbf{\textbf{0.9878}} \\
          &           & 0.1       & 28.7527/          & 26.6744/ & 30.3751/ & 25.7211/ & 26.5576/ & \textit{31.1812/} & \uline{32.9241/}  & \textbf{33.3784/}        \\
          &           &           & 0.9438            & 0.9467   & 0.9344   & 0.8904   & 0.9386   & \textit{0.9665}   & \uline{0.9767}    & \textbf{\textbf{0.9786}} \\
          & Uniform   & 0.05      & 32.9657/          & 28.4326/ & 32.7938/ & 28.5209/ & 28.8178/ & \uline{36.8639/}  & \textit{35.7976/} & \textbf{38.0102/}        \\
          &           &           & 0.9712            & 0.9595   & 0.9469   & 0.9476   & 0.9707   & \uline{0.9848}    & \textit{0.9703}   & \textbf{\textbf{0.9883}} \\
          &           & 0.1       & 30.8408/          & 27.9889/ & 31.8994/ & 27.4338/ & 28.8178/ & \textit{33.2566/} & \uline{34.6930/}  & \textbf{34.8856/}        \\
          &           &           & 0.9588            & 0.9526   & 0.9367   & 0.9370   & 0.9707   & \textit{0.9737}   & \uline{0.9809}    & \textbf{\textbf{0.9814}} \\
CVC09\_50 & Gaussian  & 0.05      & 34.7356/          & 30.0700/ & 35.4296/ & 30.2401/ & 28.7546/ & \textit{35.9095/} & \uline{36.1112/}  & \textbf{38.8772/}        \\
          &           &           & 0.9706            & 0.9592   & 0.9559   & 0.9583   & 0.9630   & \textit{0.9748}   & \uline{0.9818}    & \textbf{\textbf{0.9884}} \\
          &           & 0.1       & 30.7202/          & 27.7588/ & 29.8125/ & 26.2020/ & 25.4517/ & \textit{33.3100/} & \uline{35.7263/}  & \textbf{36.8862/}        \\
          &           &           & 0.9337            & 0.9481   & 0.8753   & 0.8816   & 0.9071   & \textit{0.9618}   & \uline{0.9816}    & \textbf{\textbf{0.9839}} \\
          & Uniform   & 0.05      & \textit{38.9405/} & 28.4200/ & 35.3353/ & 32.0414/ & 28.9208/ & \uline{39.7796/}  & 37.2899/          & \textbf{42.2163/}        \\
          &           &           & \textit{0.9799}   & 0.9517   & 0.9566   & 0.9663   & 0.9627   & \uline{0.9800}    & 0.9538            & \textbf{\textbf{0.9889}} \\
          &           & 0.1       & 34.2054/          & 30.1000/ & 33.0394/ & 29.6936/ & 28.6072/ & \textit{36.1022/} & \uline{37.6136/}  & \textbf{38.9347/}        \\
          &           &           & 0.9596            & 0.9586   & 0.9228   & 0.9552   & 0.9595   & \textit{0.9701}   & \uline{0.9823}    & \textbf{\textbf{0.9842}} \\ \hline
\end{tblr}
\end{table*}

Correspondingly, during the process of $\mathcal{S} \rightarrow \hat{\mathcal{C}} \rightarrow \widetilde{\mathcal{S}}$, the uneven distribution of stripe noise along the horizontal direction makes it no longer appropriate to directly compute the image similarity of $\mathcal{S}$ to $\widetilde{\mathcal{S}}$.
Therefore, we leverage the directional property of the stripes to compute the gradient images in the vertical direction using $\nabla_{v}(\cdot)$ and impose constraints on them.
\begin{equation}
\begin{split}
\mathcal{L}_{cyc\_s} &= \mathbb{E}_{s\sim p(s)}[\alpha (1-MS\text{-}SSIM(\nabla_{v} (\widetilde{\mathcal{S}}), \nabla_{v} (\mathcal{S}))\\
&+(1-\alpha )G_{\sigma _{G}^{m} } \mathcal{L}_{1}(\nabla_{v} (\widetilde{\mathcal{S}}), \nabla_{v} (\mathcal{S}))] 
\end{split}
\end{equation}

\subsection{Haar wavelet background guidance module (HBGM)}
To further minimize the divergence of background details between the stripe and clean domains, we propose HBGM as an additional constraint in the process of $\mathcal{S} \rightarrow \hat{\mathcal{C}}$. 
As shown in Fig.~\ref{fig:1}, the HBGM first performs a multi-level Haar wavelet transform on the image. Then, considering the stripe noise only disrupts the horizontal gradients of the image, we set the gray value of all low-frequency and vertical subbands to zero to
remove all vertical information from the images.
Finally, the inverse Haar wavelet transform is performed to restore the image details.
Comparing the results of processing two images from different domains using HBGM, it can be observed that the stripe noise and vertical information in the background are removed without discrimination on both images, regardless of the domain they belong to.
This phenomenon inspires us to use the results of HBGM for computing the similarity between images in stripe and clean domain as a powerful background content constraint, which can be established as follows:
\begin{equation}
\begin{split}
\mathcal{L}_{H} &= \alpha (1-MS\text{-}SSIM(HBGM(\mathcal{S}),HBGM(\hat{\mathcal{C}}))\\
&+(1-\alpha )G_{\sigma _{G}^{m} } \mathcal{L}_{1}(HBGM(\mathcal{S}),HBGM(\hat{\mathcal{C}})
\end{split}
\end{equation}

\subsection{Generator and Discriminator}
\subsubsection{Generator}

As shown in Fig.~\ref{fig:3}(a), we present the multi-level wavelet U-Net (MWUNet) as the generator in our DestripeCycleGAN, with the first three stages utilized for feature encoding and the remaining four stages employed for background reconstruction. 
Given a stripy image $\mathbf{I_{S}} \in \mathbb{R}^{H \times W \times 3}$, MWUNet employs seven stages for feature extraction and image restoration to achieve end-to-end mapping from $\mathbf{I_{S}}$ to the clean image $\mathbf{I_{C}}$.
Each stage is composed of convolutional layers and densely connected residual (DCR) blocks~\cite{park2019densely} to facilitate beneficial information propagation. 
Subsequently, we use the Haar wavelet transform and inverse Haar wavelet transform to replace the conventional sampling process, enriching the structural information presentation. 

Considering the importance of multi-scale information fusion~\cite{yin2023multiscale} and long-range semantic build ~\cite{li2022mafusion} in image processing tasks, we design the group fusion block (GFB) to combine low-level details with high-level semantics in skip connections. 
The detailed structure of the GFB is illustrated in Fig.~\ref{fig:3}(b). 
First, the step convolution and interpolation are used to standardize the size of different-level features. Then, these two group features are divided into eight blocks, and different-level features at the same position are fused with each other to reconstruct four fusion features.
Next, we use the different sizes of convolution kernels to establish multi-scale representations of the above fusion features.
Finally, we employ triplet attention~\cite{misra2021rotate} to separately construct the global semantic information of multi-scale features, and utilize channel-wise concatenation and pixel-by-pixel compression to achieve the completion and aggregation of the contextual information. 
In the output stage of MWUNet, a convolutional layer and Tanh activation function are used to enhance the model representation and generate a residual image, followed by element-wise addition with the input image to obtain a clean image $\mathbf{I_{C}} \in \mathbb{R}^{H \times W \times 3}$.

In previous image reconstruction work, deep perceptual networks are used to assess the similarity of different images by calculating their feature space distance~\cite {johnson2016perceptual}.
Similarly, we employ perceptual loss $\mathcal{L}_{perc}$ to mitigate unpleasant artifacts and contrast transformations.
\begin{equation}
\mathcal{L}_{perc} = \left \| \phi_{l}(\mathcal{S})- \phi_{l}(\mathcal{\hat{C}}) \right \|_{2}^{2}
\end{equation}
where $\phi_{l}(\cdot)$ stands the $conv_{2, 3}$ layer of VGG-16~\cite{simonyan2014very} network pretrained on ImageNet~\cite{deng2009imagenet}.
Besides, we also employ the identity loss $\mathcal{L}_{iden}$, which retains the details of the source domain, to safeguard the structural integrity of the clean domain~\cite{zhu2017unpaired}. The identity loss is defined as follows:
\begin{equation}
\begin{split}
\mathcal{L}_{iden} &= \mathbb{E}_{c\sim p(c)}[\alpha (1-MS\text{-}SSIM(\mathcal{C},G_{\mathcal{S} \rightarrow \mathcal{C}}(\mathcal{C}))\\
&+(1-\alpha )G_{\sigma _{G}^{m} } \mathcal{L}_{1}(\mathcal{C},G_{\mathcal{S} \rightarrow \mathcal{C}}(\mathcal{C}))] 
\end{split}
\end{equation}

\subsubsection{Discriminator}
After generating the fake image, we employ the Markovian discriminators to distinguish it from the real image~\cite{lee2018diverse}. This discriminator uses a multiscale structure in which features at each scale pass through five convolutional layers, followed by the sigmoid activation layer.
The adversarial loss $\mathcal{L}_{adv}$ between the generator and the discriminator can represented as
\begin{equation}
\label{deqn_ex1}
\begin{split}
\mathcal{L}_{adv} &=\mathbb{E}_{c\sim p(c)}  [logD_{c} (\mathcal{C})] \\
&+\mathbb{E}_{s\sim p(s)} [log(1-D_{c}(G_{\mathcal{S} \rightarrow \mathcal{C}}(\hat{\mathcal{S}})))] \\
&+\mathbb{E}_{s\sim p(s)} [log(1-D_{c}(G_{\mathcal{S} \rightarrow \mathcal{C}}(S)))]
\end{split}
\end{equation}
in which $D_{c}$ maximizes the loss function to distinguish the generated image in both branches and the real clean images, while $G_{\mathcal{S} \rightarrow \mathcal{C}}$ minimizes the objective function to make the generated image more realistic.

\begin{figure*}[t]
    \centering
    \setlength{\abovecaptionskip}{0.cm}
    \includegraphics[width=1\textwidth]{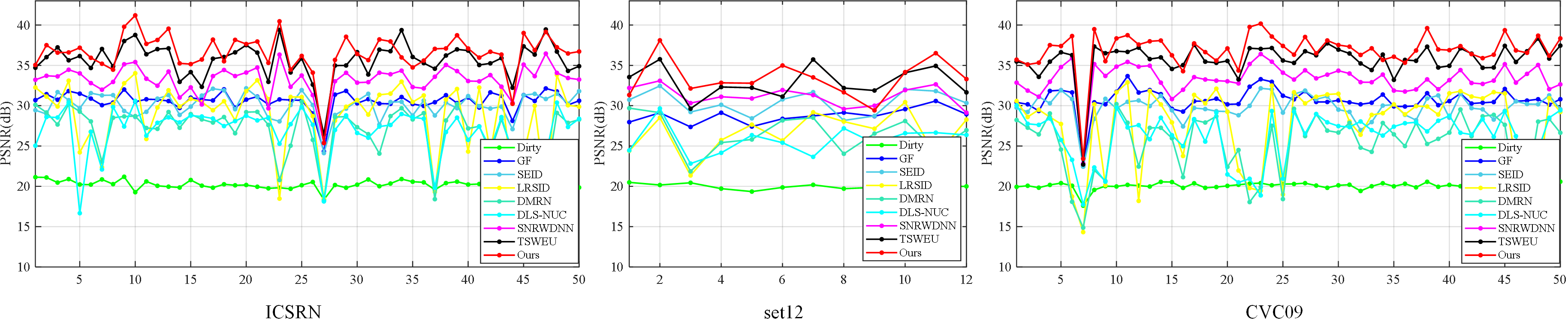}
    \caption{PSNR of different methods on DLS\_50, Set12 and CVC09\_50 with Gaussian noise of $\sigma$ = 0.1.}
    \label{fig:4_0}
\end{figure*}

\subsection{Total loss}
In this section, we combine the loss functions $\mathcal{L}_{H}$ and $\mathcal{L}_{perc}$, which are used to calculate the loss between different domains, into a single loss function denoted as $\mathcal{L}_{cross}$.
\begin{equation}
\label{formula:9}
\mathcal{L}_{cross} =  \mathcal{L}_{H} + k\mathcal{L}_{perc}
\end{equation}
and our final loss for unsupervised training of DestipingCycleGAN is presented as
\begin{equation}
\label{formula:10}
\begin{split}
\mathcal{L}_{total} &=  \lambda_{1} \mathcal{L}_{adv} + \lambda_{2}\mathcal{L}_{cyc\_s} + \lambda_{3}\mathcal{L}_{cyc\_c} \\
&+\lambda_{4}\mathcal{L}_{identity}+\lambda_{5}\mathcal{L}_{cross}
\end{split}
\end{equation}
where $k$ and $\lambda_{n}~(n=1,2,3,4,5)$ is the trade-off parameter to control the significance of the losses mentioned above.

\begin{figure*}[t]
    \centering
    \setlength{\abovecaptionskip}{0.cm}
    \includegraphics[width=0.9\textwidth]{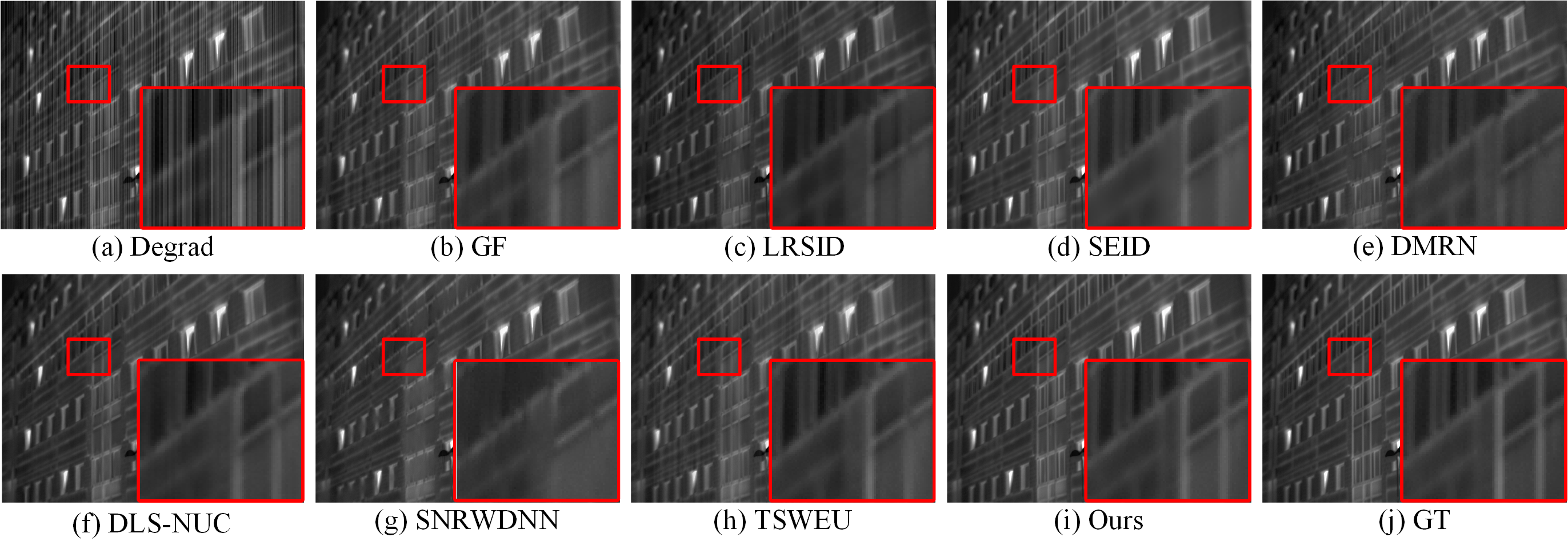}
    \caption{Comparison of destriping effects among different methods on the test image of DLS\_50~\cite{he2018single}.}
    \label{fig:4_1}
\end{figure*}

\begin{figure*}[t]
    \centering
    \setlength{\abovecaptionskip}{0.cm}
    \includegraphics[width=0.9\textwidth]{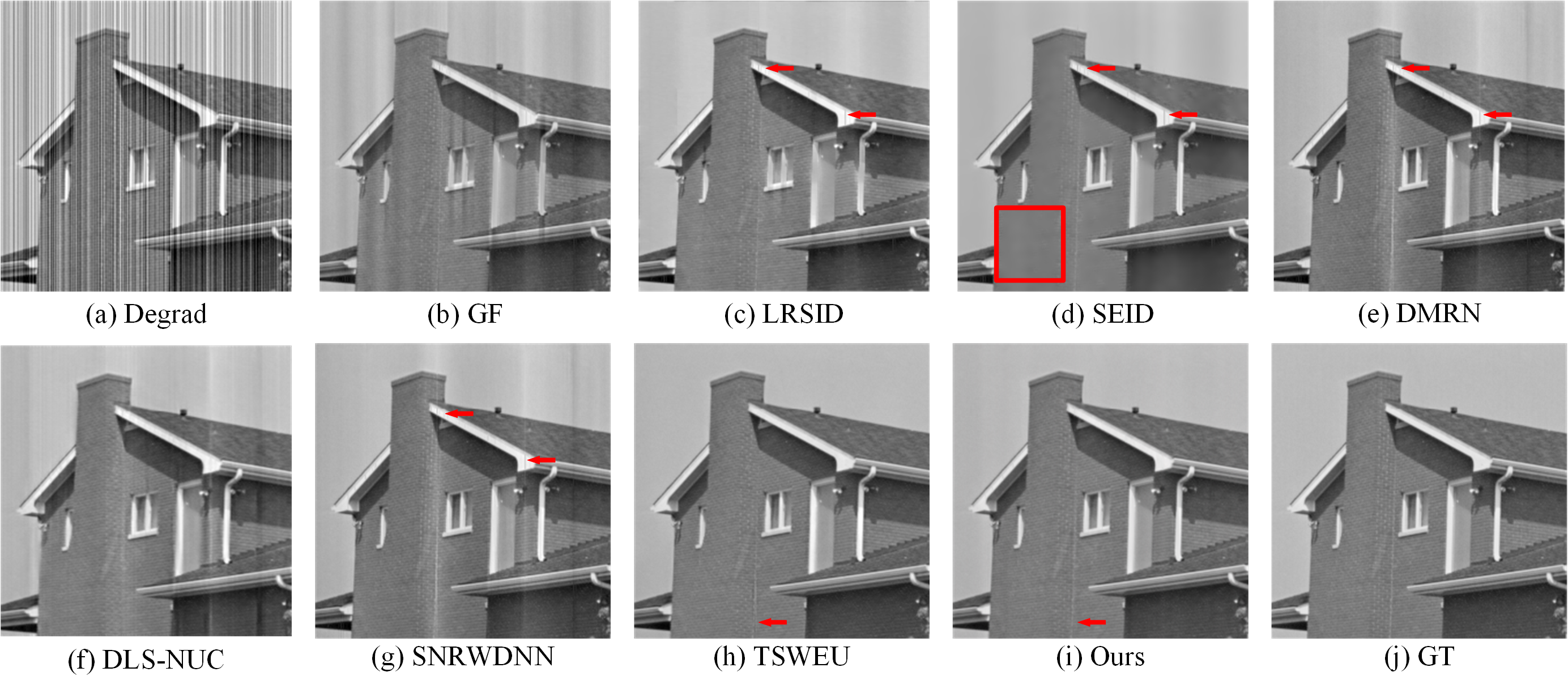}
    \caption{Comparison of destriping effects among different methods on the test image of Set12~\cite{zhang2017beyond}.
    }
    \label{fig:4_2}
\end{figure*}

\begin{figure*}[t]
    \centering
    \setlength{\abovecaptionskip}{0.cm}
    \includegraphics[width=0.9\textwidth]{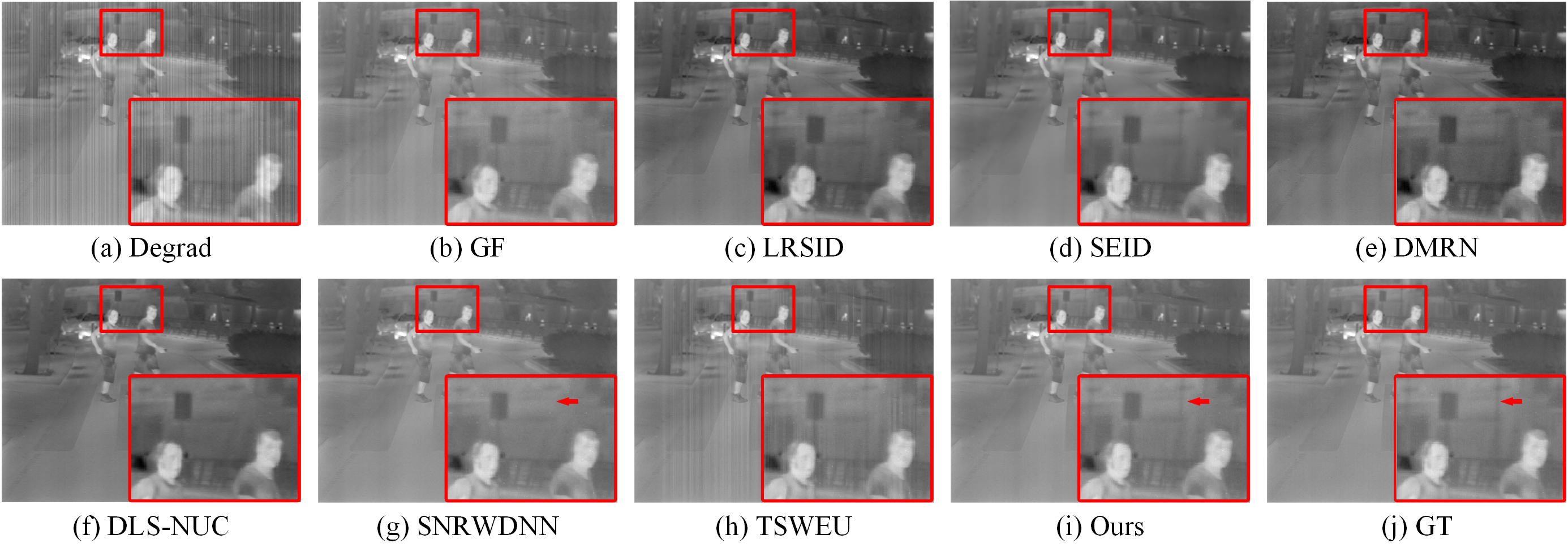}
    \caption{Comparison of destriping effects among different methods on the test image of CVC09\_50~\cite{socarras2013adapting}.
    }
    \label{fig:4_3}
\end{figure*}

\begin{figure*}[t]
    \centering
    \setlength{\abovecaptionskip}{0.cm}
    \includegraphics[width=1\textwidth]{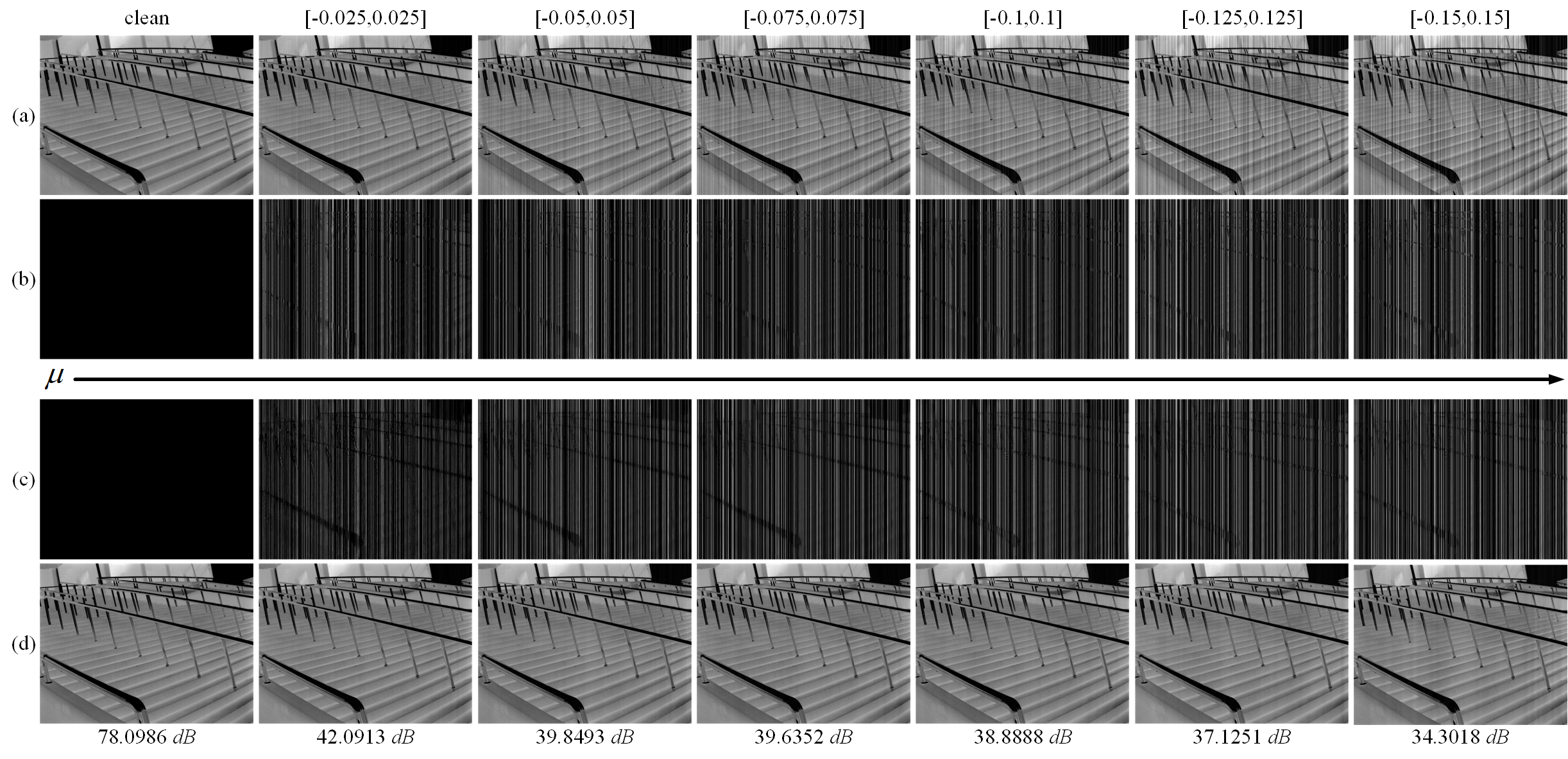}
    \caption{Performance analysis of removing stripes of different intensities on the test image of DLS\_50~\cite{he2018single}. (a) Simulated Uniform noise images with different intensities; 
    (b) Simulated stripe noises; (c) Stripe noise component removed by DestripeCycleGAN; (d) Clean background reconstructed by DestripeCycleGAN.
    }
    \label{fig:4_5}
\end{figure*}

\begin{figure*}[t]
    \centering
    \setlength{\abovecaptionskip}{0.cm}
    \includegraphics[width=1\textwidth]{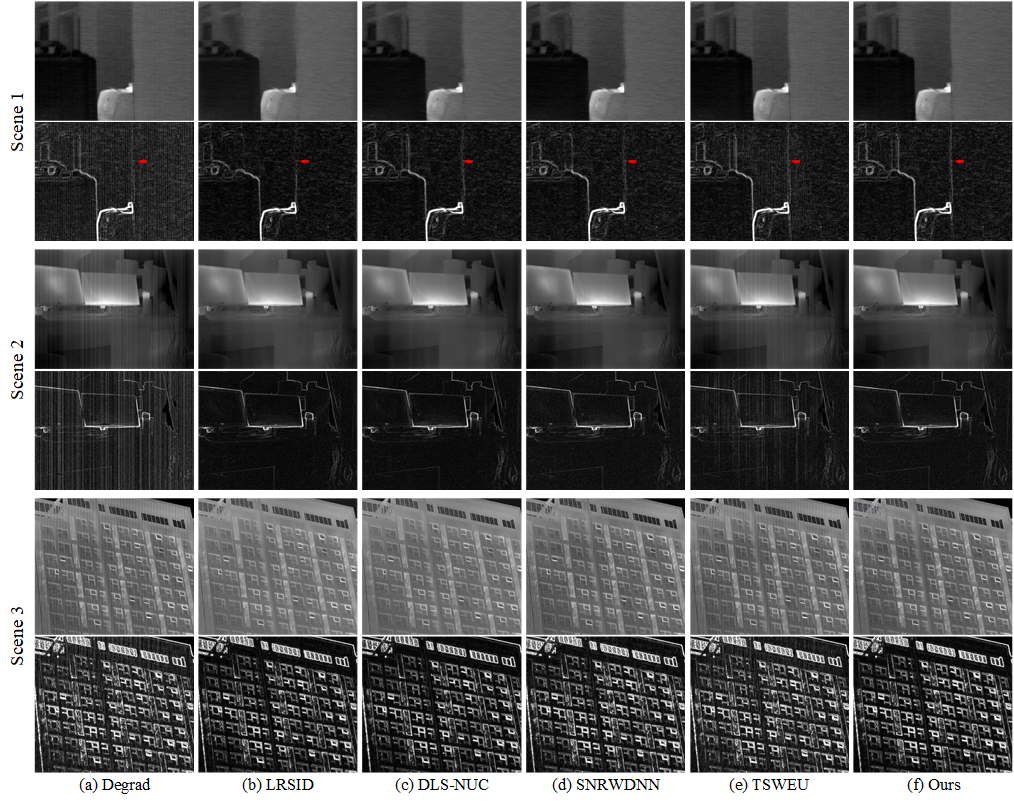}
    \caption{Destriping results of different methods on the real images. Scenes 1, 2, and 3 are from MIRE~\cite{tendero2012non}, DLS\_50~\cite{he2018single}, and our long wave infrared camera, respectively. An additional row of Sobel edges is displayed to showcase the noise level and vertical edges better.
    }
    \label{fig:5}
\end{figure*}

\section{Experiments}
\subsection{Experiment Settings}
\noindent {\textbf{Implementation Details:~}}We conducted relevant experiments on a single 2080Ti GPU. The batch size and epoch are set to 80 and 150. We optimize the model using the Adam optimizer. The cosine annealing learning strategy was used, with the initial learning rate set to 0.001. For Eq.~\ref{formula:9} and Eq.~\ref{formula:10}, we set $\lambda_{1}, \lambda_{2}, \lambda_{3}, \lambda_{4}, \lambda_{5}$ and $k$ to 1,~100,~10,~10,~10, and~0.001 empirically. 

\noindent {\textbf{Metrics:~}}Following the previous infrared image desrtiping works~\cite{kuang2018robust}, we employ peak signal-to-noise ratio (PSNR)~\cite{huang2022winnet} and structural similarity index measurement (SSIM)~\cite{wang2004image} to evaluate the effectiveness of models. We compare the DestripeCycleGAN with seven state-of-the-art methods, including GF~\cite{cao2015effective}, LRSID~\cite{chang2016remote}, SEID~\cite{song2023simultaneous}, DLS-NUC~\cite{he2018single}, DMRN~\cite{chang2019infrared}, SNRWDNN~\cite{guan2019wavelet}, and TSWEU~\cite{chang2019toward}.

\noindent {\textbf{Datasets:~}}For the training data in the clean domain, we select 300 images from DLS-NUC~\cite{he2018single}, CVC-09~\cite{socarras2013adapting}, and our mid wave infrared cooled camera. We present Uniform noise with $\mu$ taking random values in $[0.5, 0.1]$ to these 300 images and added 150 stripe images from real scenes selected from MIRE~\cite{tendero2012non}, SNRWDNN~\cite{guan2019wavelet}, CVC-09, and our long wave infrared uncooled camera. In total, 450 images are used as training data in the stripe domain. Using rotation and scaling, the training data is cropped into 64$\times$64 image patches, resulting in 25,688 images in the clean domain and 33,477 images in the stripe domain. For test data, we construct three sets, DLS\_50, CVC09\_50, and Set12~\cite{zhang2017beyond}, to widely evaluate destriping performance.

\begin{figure*}[t]
    \centering
    \setlength{\abovecaptionskip}{0.cm}
    \includegraphics[width=0.8\textwidth]{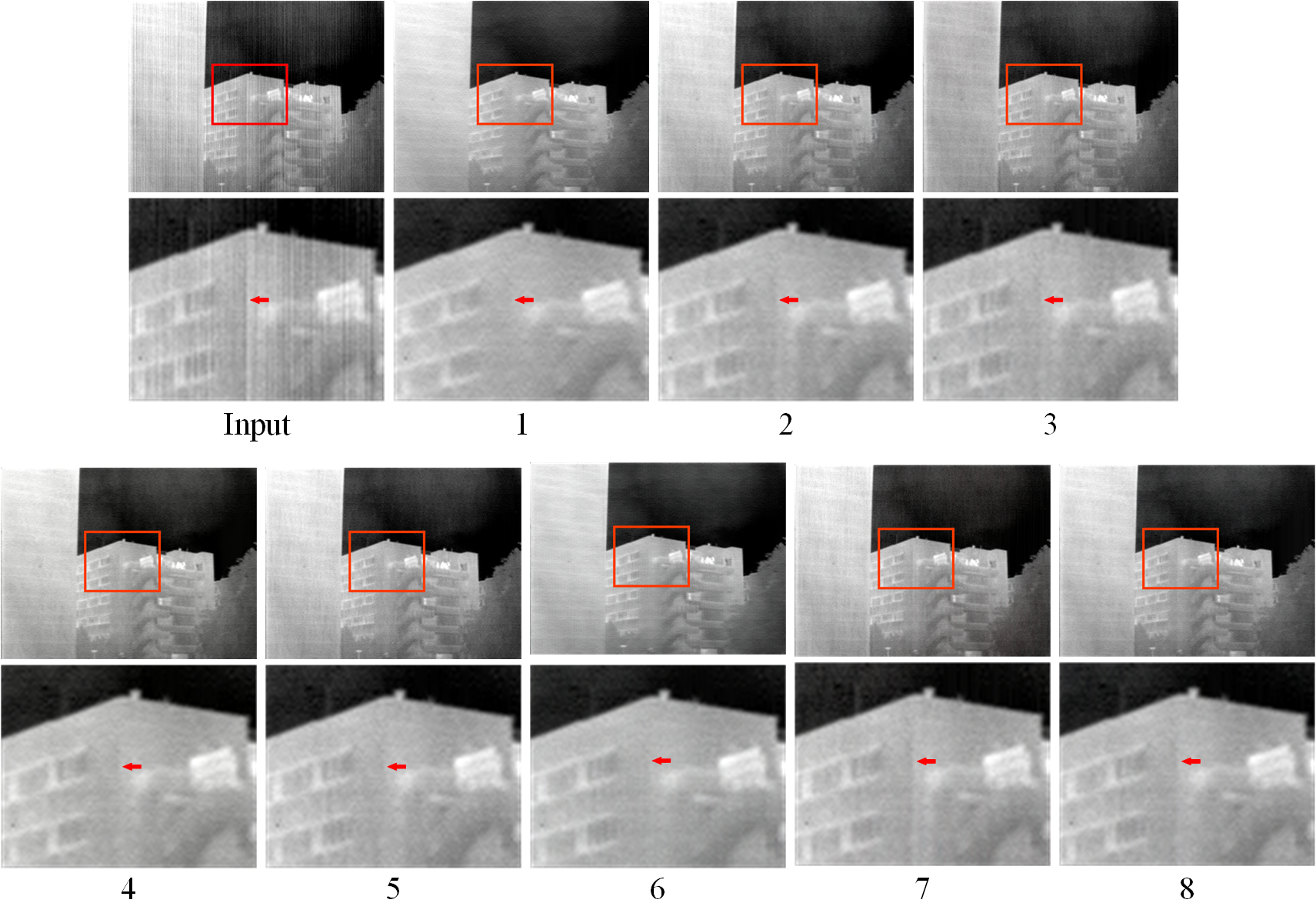}
    \caption{Comparison of destriping results on MIRE~\cite{tendero2012non} using different loss function combinations from Table ~\ref{table3}. To emphasize the details of the background, the region marked by the red box in the figure below is enlarged.
    }
    \label{fig:6}
\end{figure*}

\subsection{Destriping Results for Simulated stripes}
\noindent {\textbf{Quantitative results:~}}Table \ref{table3} 
lists the quantitative results of the various de-striping methods. We add Gaussian noise with $\sigma = 0.05$ and $\sigma = 0.1$, while Uniform noise with $\mu = 0.05$ and $\mu = 0.1$ to each test set. The best three results are shown in bold, underlined, and italicized fonts, sequentially. Our methods are superior to the state-of-the-art in all instances. For example, when considering Uniform noise on DLS\_50, the PSNR values of our method are 3.10 dB and 4.68 dB higher than SNRWDNN and TSWEU, respectively. Note that although the SGM in the proposed framework consistently generates Gaussian noise during the training phase, our DestripeCycleGAN exhibits optimal performance in both noise modes.

We add Gaussian stripe noise with $\sigma = 0.1$ to DLS\_50, Set12, and CVC09\_50 datasets and present the PSNR index curves of various methods on these three test sets in Fig.~\ref{fig:4_0}. 
It is evident that our DestripeCycleGAN and TSWEU produce the most satisfactory results compared to other methods, and our method outperforms TSWEU on the majority of images. 
Furthermore, despite observed fluctuations in the curves corresponding to specific images, our method consistently maintains superiority over other destriping methods.

\noindent {\textbf{{Visual Performance:~}}To further evaluate the effectiveness of the DestripeCycleGAN, we selected representative images in three datasets and visualized the results of all methods on the images. 

Fig.~\ref{fig:4_1} shows a representative image from DLS\_50 that is corrupted by a Uniform noise with $\mu = 0.1$ and has distinct vertical edges. From the sub-images (b), (c), and (h), we observe significant stripe noise residuals, while SEID and DLS-NUC contain slight noise residuals as shown in (d) and (f). 
Although DMRN and SNRWDNN successfully eliminate stripe noise, they unintentionally erase the window's vertical edges that overlap with the stripe noise. Our method is the only one that can effectively eliminate stripe noise while maintaining vertical edges. 
This is achieved through the SGM to compensate for the semantic difference between vertical edges of windows and stripe noise as well as the MWUnet to minimize the loss of background information during feature extraction.

Fig.~\ref{fig:4_2} shows the visualization results on the test image of Set12, which is disturbed by Gaussian noise with $\sigma = 0.1$. 
In this scene, there are obvious noise residues as shown in (b), (e), (f), and (g).
Additionally, LRSID, SEID, DLS-NUC, and SNRWDNN show minor noise residues as indicated by the infrared arrow in (c), (d), (e), and (g), with SEID causing excessive image smoothing in the region marked by the red rectangle.
Both TSWEU and our method demonstrate satisfactory results in that scene, but our method exhibited superior performance in preserving intricate details in the corner of the wall. 
This effectively demonstrates the superiority of the MWUnet and HBGM co-optimization model in preserving background details and image texture.

Fig.~\ref{fig:4_3} shows an image from CVC09\_50 degraded by a Uniform noise with $\mu =0.05$. The denoising ability of GF and TSWEU remains poor under weaker Uniform noise perturbation, and LRSID, SEID, DMRN, and DLS-NUC result in noticeable brightness variations as exhibited in (c), (d), (e), and (f). 
Although SNRWDNN can remove most of the noise, we can observe that the method removes both the utility poles and the stripe noise from the background. 
Thanks to DestripeCycleGAN's robust domain translation strategy, our method successfully eliminates stripe noise in this scene while maintaining the consistent overall brightness of the image. 

\noindent {\textbf{Anti-noise Experiment:~}}The effectiveness of the proposed method in removing stripe noise of different intensities is demonstrated in Fig.~\ref{fig:4_5}. Despite the multiplicative noise, our algorithm can still identify the stripe noise that overlaps with the ground truth (GT) and successfully obliterate it. 
Nevertheless, as the intensity of the stripe noise increases, a minor residual noise may be left in the results. This is because the maximum intensity of simulated noise is set to 0.12 in SGM and 0.1 during the training, which leads to a slight degradation in the model's ability to remove intense stripe noise effectively. However, DestripeCycleGAN can still remove most of the stripe noise in both the $\mu = 0.125$ and $\mu = 0.15$ cases, which further illustrates its robust destriping capability.

\subsection{Results on Real Noisy Images}
To further assess the performance of the proposed method in various noise modes, we conducted tests on authentic infrared stripe noise images using DestripeGAN, as depicted in Fig~\ref{fig:5}. The results indicate that DestripeCycleGAN can remove stripe noise completely and excels at recovering the vertical details of the image. For instance, in scene 1, only our method can render the vertical edges of the wall entirely. However, in some cases, other methods may fail, e.g., TSWEU, which performs well on the simulated dataset, exhibits significant stripe noise residuals in Scenes 2 and 3. We attribute this to the fact that TSWEU is a network based on stripe noise modeling, which results in poor performance when dealing with noise from different sources. The model-driven method LRSID shows competitive performance in the three scenarios, thanks to its modeling of the stripes' low-rank properties that help remove the weak real noise. In contrast, the proposed method is robust and can adapt to various noise patterns across different scenarios.

\subsection{Ablation Study and Discussion}
In this section, we study the effectiveness of each term in this paper on the DLS\_50 for a Uniform-distributed stripe noise of $\mu =0.05$. 
We first validate the performance of the DestripeCycleGAN framework and MWUNet generator. 
Therefore, we construct four different combinations based on the traditional CycleGAN framework and U-Net generator.

\begin{table}[h]
\centering
\caption{Ablation Study for DestripeCycleGAN and MWUNet}
\label{table2}
\renewcommand{\arraystretch}{1.3}
\setlength\tabcolsep{0.35mm}
\begin{tabular}{ccccc}
\hline
Metrics       & \begin{tabular}[c]{@{}c@{}}CycleGAN\\ +U-Net\end{tabular} & \begin{tabular}[c]{@{}c@{}}DestripeCycleGAN\\ +U-Net\end{tabular} & \begin{tabular}[c]{@{}c@{}}CycleGAN\\ +MWUNet\end{tabular} & \begin{tabular}[c]{@{}c@{}}DestripeCycleGAN\\ +MWUNet\end{tabular} \\ \hline
PSNR          & 39.0334  & 40.3455  & 41.0441  & \textbf{42.9974}  \\
SSIM          & 0.9686   & 0.9809   & 0.9765   & \textbf{0.9886}   \\
Params (M) & 20.03    & 	\textbf{10.01}   & 31.02  & 15.51  \\ 
FLOPS (G)      & 36.60    & 	\textbf{18.00}   &75.85   & 37.93  \\ \hline
\end{tabular}
\end{table}

The experimental results are shown in Table~\ref{table2}. It can be observed that with the same generator, the proposed DestripeCycleGAN framework substantially improves the image destriping performance compared to the original CycleGAN while reducing model parameters and computational complexity by half.
This advantage is attributed to our framework replacing the auxiliary generator in the original CycleGAN with SGM, which not only reduces the parameter but also makes the generated stripe noise align more closely with its structural property distribution.
Furthermore, under different destriping frameworks, MWUNet achieves higher PSNR values compared to U-Net by 2.01 and 2.65 dB, respectively. That illustrates the overall effectiveness of MWUNet.
Finally, comparing DestripeCycleGAN+MWUNet and CycleGAN+U-Net, we find that the proposed method can greatly improve algorithm performance while reducing model parameters, almost without altering computational complexity.

\begin{table}[t]
\centering
\footnotesize
\caption{Destriping results (PSNR/SSIM) with different combinations of loss function}
\label{table3}
\renewcommand{\arraystretch}{1.3}
\setlength\tabcolsep{1mm}
\begin{tabular}{ccccccccc}
\hline
Case & $\mathcal{L}_{adv}$ & $\mathcal{L}_{cyc}\_c$ & $\mathcal{L}_{cyc}\_s$ & $\mathcal{L}_{iden}$ & $\mathcal{L}_{cross}$ & $\mathcal{L}_1$ & PSNR                              & SSIM                             \\ \hline
1    & $\checkmark$        &                        &                        &                      & $\checkmark$          &                 & 38.6906                           & 0.9734                           \\
2    & $\checkmark$        & $\checkmark$           &                        &                      & $\checkmark$          &                 & 42.0782                           & 0.9852                           \\
3    & $\checkmark$        & $\checkmark$           & $\checkmark$           &                      & $\checkmark$          &                 & 42.4835                           & 0.9874                           \\
4    & $\checkmark$        &                        &                        & $\checkmark$         & $\checkmark$          &                 & 39.7687                           & 0.9730                           \\
5    & $\checkmark$        & $\checkmark$           &                        & $\checkmark$         & $\checkmark$          &                 & 42.6943                           & 0.9870                           \\
6    & $\checkmark$        & $\checkmark$           & $\checkmark$           & $\checkmark$         &                       &                 & 42.2685                           & 0.9877                           \\
7    & $\checkmark$        & $\checkmark$           & $\checkmark$           & $\checkmark$         & $\checkmark$          & $\checkmark$    & 41.5716                           & 0.9846                           \\
8    & $\checkmark$        & $\checkmark$           & $\checkmark$           & $\checkmark$         & $\checkmark$          &                 & \textbf{42.9974} & \textbf{0.9886} \\ \hline
\end{tabular}
\end{table}

\subsubsection{Loss Function}
In Table~\ref{table3}, we construct different combinations of loss functions as cases 1-7 to validate the effectiveness of the unsupervised constraints, and the corresponding visual performance is shown in Fig.~\ref{fig:6}.
The results of case 1 indicate that the model can learn the mapping from the striped domain to the clean domain without the identity loss $\mathcal{L}_{iden}$ and the cycle-consistency loss $\mathcal{L}_{cyc}$, even achieves performance comparable to other deep learning methods. 
By comparing cases 2 and 3 along with 5 and 8, our proposed cycle-consistent loss $\mathcal{L}_{cyc\_s}$ in the stripe-to-stripe domain can be used to preserve the image details and thus improve the imaging quality by guiding the directional properties of the stripes. 
The contrasts between cases 1 and 4, 2 and 5, then 3 and 8 illustrate that the identity loss $\mathcal{L}_{iden}$ helps to remove the weak stripe noise residuals and retains the luminance and contrast information of the clean image to enhance the visual effect significantly. 
Case 6 demonstrates that both the cross-domain loss $\mathcal{L}_{cross}$ play an essential role in preserving the vertical edges of the background, ultimately leading to significant improvements in image quality. 
Case 7 represents replacing all the loss functions in the computation of pixel-level losses in the image to $\mathcal{L}_1$, and the results show a relatively significant decrease in the model performance.

\begin{table}[t]
\centering
\footnotesize
\caption{Objective results (PSNR/SSIM) of removing different components of DestripeCycleGAN}
\label{table4}
\renewcommand{\arraystretch}{1.3}
\begin{tabular}{c|cccc}
\hline
          & \multicolumn{4}{c}{PSNR}                                                                                                                                                       \\ \hline
Methods   & \begin{tabular}[c]{@{}c@{}}Ours\\ $-$~\{$\mathcal{L}_{prev}, \mathcal{L}_H$\}\end{tabular} & \begin{tabular}[c]{@{}c@{}}Ours\\ $-$~\{$\mathcal{L}_{prev}$\}\end{tabular} & \begin{tabular}[c]{@{}c@{}}Ours\\ $-$~\{$\mathcal{L}_H$\}\end{tabular} & \textbf{Ours}    \\ \hline
DLS\_50   & 42.2685                                                    & 42.7628                                            & 42.6975                                            & \textbf{42.9974} \\
Set12     & 36.4178                                                    & 37.7866                                            & 37.3903                                            & \textbf{38.0102} \\
CVC09\_50 & 41.7169                                                    & 42.0213                                            & 41.9570                                            & \textbf{42.2163} \\ \hline
          & \multicolumn{4}{c}{SSIM}                                                                                                                                                       \\ \hline
DLS\_50   & 0.9877                                                     & 0.9881                                             & 0.9882                                             & \textbf{0.9886}  \\
Set12     & 0.9808                                                     & 0.9880                                             & 0.9870                                             & \textbf{0.9883}  \\
CVC09\_50 & 0.9868                                                     & 0.9876                                             & 0.9870                                             & \textbf{0.9889}  \\ \hline
\end{tabular}
\end{table}

\subsubsection{Haar wavelet background guidance module (HBGM)}
To verify the effectiveness of the proposed HBGM in cross-domain constraints, we select the proposed DestripeCycleGAN as a baseline and design three ablation schedules in Table~\ref{table4}.
Due to space limitations, we use the abbreviation \textbf{``Ours''} to refer to the complete DestripeCycleGAN.

\begin{itemize} 
\item \textbf{Ours~$\text{-}$~}\bm{$\{\mathcal{L}_{prev}, \mathcal{L}_H\}$}: Benchmark scheme without the loss of $\mathcal{L}_{prev}$ and $\mathcal{L}_H$.
\item \textbf{Ours~\text{-}~}\bm{$\mathcal{L}_{prev}$}: Remove perceptual loss $\mathcal{L}_{prev}$.
\item \textbf{Ours~\text{-}~}\bm{$\mathcal{L}_H$}: Remove the loss from the Haar wavelet background guidance module.
\end{itemize} 

\begin{figure}[b]
    \centering
    \setlength{\abovecaptionskip}{0.cm}
    \includegraphics[width=0.45\textwidth]{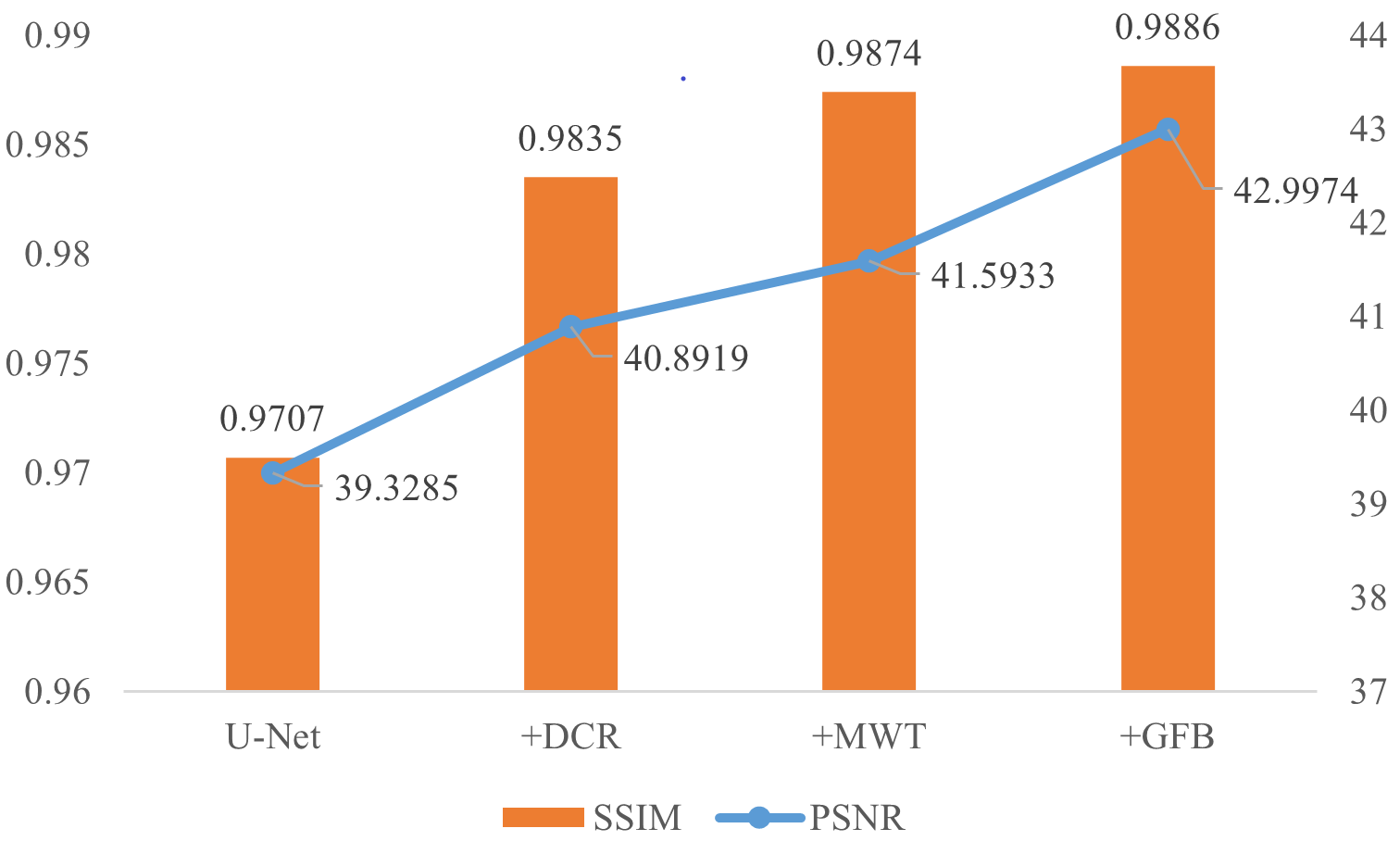}
    \caption{Ablation study of the proposed MWUNet.
    }
    \label{fig:6_1}
\end{figure}

It can be shown that the complete DestripeCycleGAN achieves 0.73/0.23/0.30 dB, 1.59/0.22/0.62 dB, and 0.50/0.20/26 dB PSNR gain over three ablated baselines on DLS\_50, Set12, CVC09\_50 respectively. 
Moreover, by comparing \textbf{Ours~\text{-}~}\bm{$\mathcal{L}_{prev}$} and \textbf{Ours~\text{-}~}\bm{$\mathcal{L}_H$}, we can find that HBGM has a slight advantage in cross-domain constraints. This illustrates that HBGM can more precisely evaluate the similarity between images under the interference of stripe noise.

\subsubsection{Multi-level Wavelet U-Net (MWUNet)}
Fig.~\ref{fig:6_1} illustrates the effectiveness of each module within our MWUNet.
We can see that the performance of the network shows a gradual improvement with the incorporation of additional modules. 
More precisely, the DCR block enables the dissemination of information in local areas; the multi-level wavelet transform (MWT) helps reduce information loss during reconstruction. Moreover, GFB effectively integrates multi-scale context and enhances feature representation.

\begin{table}[t]
\centering
\footnotesize
\caption{Destriping results (PSNR/SSIM) with different data paradigm}
\label{table5}
\renewcommand{\arraystretch}{1.1}
\setlength\tabcolsep{1.8mm}
\begin{tabular}{cccccc}
\hline
Metrics    & Branch  & Pair & +Real(35\%) & +Real(52\%) & All real \\ \hline
PSNR & 42.7157  & \textbf{43.2821}  & 42.9974  & 42.9100  & 41.4444      \\
SSIM & 0.9893  & \textbf{0.9901}  & 0.9886  & 0.9867  & 0.9841     \\ \hline
\end{tabular}
\end{table}

\begin{figure}[t]
    \centering
    \setlength{\abovecaptionskip}{0.cm}
    \includegraphics[width=0.45\textwidth]{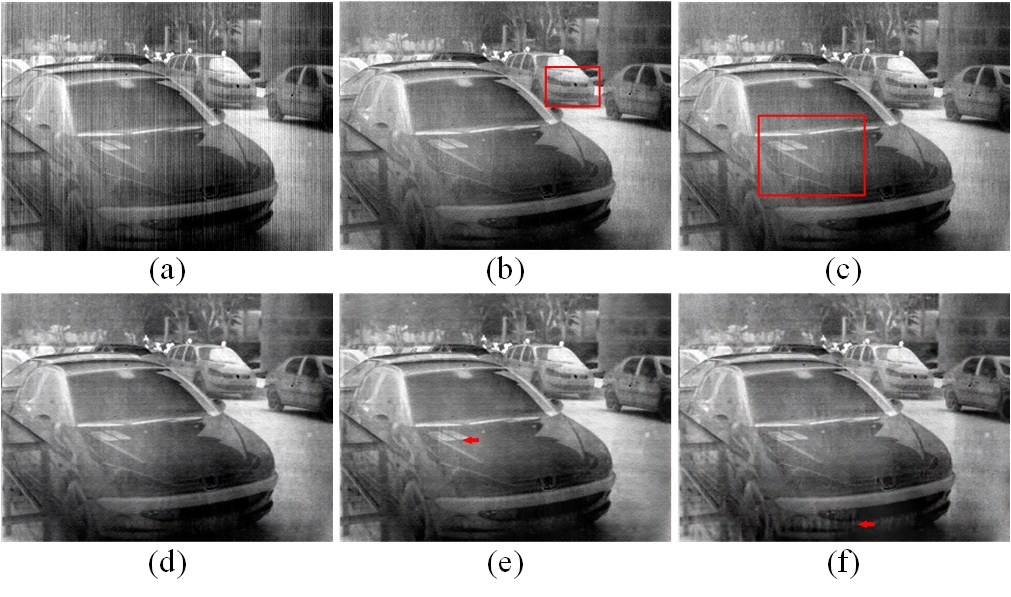}
    \caption{Destriping comparison of different data paradigms on MIRE~\cite{tendero2012non}. From (a) to (f) are the results of Input, Branch, Pair, +Real(35\%), +Real(52\%), and All real, respectively.
    }
    \label{fig:7}
\end{figure}

\subsubsection{Data paradigm}
As DestripeCycleGAN is an unsupervised model, we also aimed to examine its performance using various data pairing paradigms. 
The quantitative results are presented in Table~\ref{table5} and corresponding visual representations in the real scene are shown in Fig.~\ref{fig:7}.
\textbf{Branch} stands for training using only branch $\mathcal{C} \rightarrow \hat{\mathcal{S}} \rightarrow \widetilde{\mathcal{C}}$ and constraining using only MS-SSIM mixing $\mathcal{L}_{1}$ loss, which makes the method fully supervised. We observe that its performance is not as good as the DestripeCycleGAN framework both in simulated stripes and real vision. \textbf{Pair} stands for training with fully paired data in the proposed framework, which demonstrates high metrics on simulated stripes but exhibits significant noise residuals in real scenarios. The percentages in \textbf{+Real(35\%)} and \textbf{+Real(52\%)} represent real patches weighting in clean patches. It shows that introducing a part of real stripe images can serve as an extra to the data, and will enhance general performance. When utilizing all real data, significant disparities in image distribution of \textbf{All real} will lead to a decline in the model.

\section{Conclusion}
In this paper, we present a DestripeCycleGAN framework for unsupervised infrared image destriping.
The key point is replacing the auxiliary generator of primitive CycleGAN with the stripe generation model, allowing for more precise modeling of the structural characterization of stripe noises.
Further, the Haar wavelet background guidance module and a novel generator called multi-level wavelet U-Net are designed to collaboratively optimize the proposed DestripeCycleGAN model to recover background information. 
Comprehensive experiments demonstrate that our method outperforms existing methods on both synthetic and real datasets. 
In the future, we plan to validate the effectiveness of this unsupervised destriping paradigm for downstream tasks in real-world scenarios.

\ifCLASSOPTIONcaptionsoff
  \newpage
\fi

\bibliographystyle{IEEEtran}
\small\bibliography{IEEEabrv,reference}

\begin{thebibliography}{10}
\providecommand{\url}[1]{#1}
\csname url@samestyle\endcsname
\providecommand{\newblock}{\relax}
\providecommand{\bibinfo}[2]{#2}
\providecommand{\BIBentrySTDinterwordspacing}{\spaceskip=0pt\relax}
\providecommand{\BIBentryALTinterwordstretchfactor}{4}
\providecommand{\BIBentryALTinterwordspacing}{\spaceskip=\fontdimen2\font plus
\BIBentryALTinterwordstretchfactor\fontdimen3\font minus \fontdimen4\font\relax}
\providecommand{\BIBforeignlanguage}[2]{{%
\expandafter\ifx\csname l@#1\endcsname\relax
\typeout{** WARNING: IEEEtran.bst: No hyphenation pattern has been}%
\typeout{** loaded for the language `#1'. Using the pattern for}%
\typeout{** the default language instead.}%
\else
\language=\csname l@#1\endcsname
\fi
#2}}
\providecommand{\BIBdecl}{\relax}
\BIBdecl

\bibitem{aizat2023comprehensive}
M.~Aizat, N.~Qistina, and W.~Rahiman, ``A {C}omprehensive {R}eview of {R}ecent {A}dvances in {A}utomated {G}uided {V}ehicle {T}echnologies: {D}ynamic {O}bstacle {A}voidance in {C}omplex {E}nvironment {T}oward {A}utonomous {C}apability,'' \emph{IEEE Trans. Instrum. Meas.}, 2023.

\bibitem{wu2023mtu}
T.~Wu, B.~Li, Y.~Luo, Y.~Wang, C.~Xiao, T.~Liu, J.~Yang, W.~An, and Y.~Guo, ``M{T}{U}-{N}et: Multilevel {T}ransunet for {S}pace-{B}ased {I}nfrared {T}iny {S}hip {D}etection,'' \emph{IEEE Trans. Geosci. Remote Sens.}, vol.~61, pp. 1--15, 2023.

\bibitem{fang2022infrared}
H.~Fang, L.~Ding, L.~Wang, Y.~Chang, L.~Yan, and J.~Han, ``Infrared small uav target detection based on depthwise separable residual dense network and multiscale feature fusion,'' \emph{IEEE Trans. Instrum. Meas.}, vol.~71, pp. 1--20, 2022.

\bibitem{yuan2024sctransnet}
S.~Yuan, H.~Qin, X.~Yan, N.~AKhtar, and A.~Mian, ``S{CT}ransnet: {S}patial-channel {C}ross {T}ransformer {N}etwork for {I}nfrared {S}mall {T}arget {D}etection,'' \emph{arXiv preprint arXiv:2401.15583}, 2024.

\bibitem{Badrinarayanan2017Segnet}
V.~Badrinarayanan, A.~Kendall, and R.~Cipolla, ``Segnet: A deep convolutional encoder-decoder architecture for image segmentation,'' \emph{IEEE Trans. Pattern Anal. Mach. Intell.}, vol.~39, no.~12, pp. 2481--2495, 2017.

\bibitem{cao2015effective}
Y.~Cao, M.~Y. Yang, and C.-L. Tisse, ``Effective strip noise removal for low-textured infrared images based on 1-d guided filtering,'' \emph{IEEE Trans. Circuits Syst. Video Technol.}, vol.~26, no.~12, pp. 2176--2188, 2015.

\bibitem{zeng2019fourier}
Q.~Zeng, H.~Qin, X.~Yan, and H.~Zhou, ``Fourier spectrum guidance for stripe noise removal in thermal infrared imagery,'' \emph{IEEE Geosci. Remote Sens. Lett.}, vol.~17, no.~6, pp. 1072--1076, 2019.

\bibitem{zeng2020fourier}
Q.~Zeng, H.~Qin, X.~Yan, and T.~Yang, ``Fourier domain anomaly detection and spectral fusion for stripe noise removal of {TIR} imagery,'' \emph{Remote Sens.}, vol.~12, no.~22, p. 3714, 2020.

\bibitem{tendero2012non}
Y.~Tendero, S.~Landeau, and J.~Gilles, ``Non-uniformity correction of infrared images by midway equalization,'' \emph{Image Process. Line}, vol.~2, pp. 134--146, 2012.

\bibitem{horn1979destriping}
B.~K. Horn and R.~J. Woodham, ``Destriping landsat {MSS} images by histogram modification,'' \emph{Computer Graphics and Image Process.}, vol.~10, no.~1, pp. 69--83, 1979.

\bibitem{wegener1990destriping}
M.~Wegener, ``Destriping multiple sensor imagery by improved histogram matching,'' \emph{Int. J. Remote Sens.}, vol.~11, no.~5, pp. 859--875, 1990.

\bibitem{chang2016remote}
Y.~Chang, L.~Yan, T.~Wu, and S.~Zhong, ``Remote sensing image stripe noise removal: From image decomposition perspective,'' \emph{IEEE Trans. Geosci. Remote Sens.}, vol.~54, no.~12, pp. 7018--7031, 2016.

\bibitem{song2023simultaneous}
L.~Song and H.~Huang, ``Simultaneous {D}estriping and {I}mage {D}enoising {U}sing a {N}onparametric {M}odel {W}ith the {EM} {A}lgorithm,'' \emph{IEEE Trans. Image Process.}, vol.~32, pp. 1065--1077, 2023.

\bibitem{he2023fspnp}
Y.~He, C.~Zhang, B.~Zhang, and Z.~Chen, ``{FSP}n{P}: Plug-and-{P}lay {F}requency-{S}patial {D}omain {H}ybrid {D}enoiser for {T}hermal {I}nfrared {I}mage,'' \emph{IEEE Trans. Geosci. Remote Sens.}, 2023.

\bibitem{kuang2017single}
X.~Kuang, X.~Sui, Y.~Liu, Q.~Chen, and G.~Guohua, ``Single infrared image optical noise removal using a deep convolutional neural network,'' \emph{IEEE Photon. J.}, vol.~10, no.~2, pp. 1--15, 2017.

\bibitem{xiao2018removing}
P.~Xiao, Y.~Guo, and P.~Zhuang, ``Removing stripe noise from infrared cloud images via deep convolutional networks,'' \emph{IEEE Photon. J.}, vol.~10, no.~4, pp. 1--14, 2018.

\bibitem{he2018single}
Z.~He, Y.~Cao, Y.~Dong, J.~Yang, Y.~Cao, and C.-L. Tisse, ``Single-image-based nonuniformity correction of uncooled long-wave infrared detectors: A deep-learning approach,'' \emph{Appl. Opt.}, vol.~57, no.~18, pp. D155--D164, 2018.

\bibitem{li2021Non}
T.~Li, Y.~Zhao, Y.~Li, and G.~Zhou, ``Non-uniformity correction of infrared images based on improved cnn with long-short connections,'' \emph{IEEE Photon. J.}, vol.~13, no.~3, pp. 1--13, 2021.

\bibitem{jong2020dual}
J.-H. Lee and Y.~M. Ro, ``Dual-{B}ranch {S}tructured {D}e-striping {C}onvolution {N}etwork {U}sing {P}arametric {N}oise {M}odel,'' \emph{IEEE Access}, vol.~8, pp. 155\,519--155\,528, 2020.

\bibitem{chang2019infrared}
Y.~Chang, L.~Yan, L.~Liu, H.~Fang, and S.~Zhong, ``Infrared aerothermal nonuniform correction via deep multiscale residual network,'' \emph{IEEE Geosci. Remote Sens. Lett.}, vol.~16, no.~7, pp. 1120--1124, 2019.

\bibitem{xu2022single}
K.~Xu, Y.~Zhao, F.~Li, and W.~Xiang, ``Single infrared image stripe removal via deep multi-scale dense connection convolutional neural network,'' \emph{Infr. Phys. Technol.}, vol. 121, p. 104008, 2022.

\bibitem{guan2020fixed}
J.~Guan, R.~Lai, A.~Xiong, Z.~Liu, and L.~Gu, ``Fixed pattern noise reduction for infrared images based on cascade residual attention {CNN},'' \emph{Neurocomputing}, vol. 377, no.~15, pp. 301--313, 2020.

\bibitem{ding2022single}
D.~Ding, Y.~Li, P.~Zhao, K.~Li, S.~Jiang, and Y.~Liu, ``Single {I}nfrared {I}mage {S}tripe {R}emoval via {R}esidual {A}ttention {N}etwork,'' \emph{Sensors}, vol.~22, no.~22, p. 8734, 2022.

\bibitem{li2023progressive}
J.~Li, J.~Zhang, J.~Han, C.~Yan, and D.~Zeng, ``Progressive {R}ecurrent {N}eural {N}etwork for {M}ultispectral {R}emote {S}ensing {I}mage {D}estriping,'' \emph{IEEE Trans. Geosci. Remote Sens.}, 2023.

\bibitem{yuan2024arcnet}
S.~Yuan, H.~Qin, X.~Yan, N.~Akhtar, S.~Yang, and S.~Yang, ``A{RCN}et: An {A}symmetric {R}esidual {W}avelet {C}olumn {C}orrection {N}etwork for {I}nfrared {I}mage {D}estriping,'' \emph{arXiv preprint arXiv:2401.15578}, 2024.

\bibitem{wang2022noise}
T.~Wang, Q.~Yin, F.~Cao, M.~Li, Z.~Lin, and W.~An, ``Noise {P}arameter {E}stimation {T}wo-{S}tage {N}etwork for {S}ingle {I}nfrared {D}im {S}mall {T}arget {I}mage {D}estriping,'' \emph{Remote Sens.}, vol.~14, no.~19, p. 5056, 2022.

\bibitem{cao2022robust}
S.~Cao, H.~Fang, L.~Chen, W.~Zhang, Y.~Chang, and L.~Yan, ``Robust blind deblurring under stripe noise for remote sensing images,'' \emph{IEEE Trans. Geosci. Remote Sens.}, vol.~60, pp. 1--17, 2022.

\bibitem{munch2009stripe}
B.~M{\"u}nch, P.~Trtik, F.~Marone, and M.~Stampanoni, ``Stripe and ring artifact removal with combined wavelet—fourier filtering,'' \emph{Opt. Express}, vol.~17, no.~10, pp. 8567--8591, 2009.

\bibitem{zhong2020partial}
J.~Zhong, X.~Bi, Q.~Shu, M.~Chen, D.~Zhou, and D.~Zhang, ``Partial discharge signal denoising based on singular value decomposition and empirical wavelet transform,'' \emph{IEEE Trans. Instrum. Meas.}, vol.~69, no.~11, pp. 8866--8873, 2020.

\bibitem{chui1992introduction}
C.~K. Chui, \emph{An introduction to wavelets}.\hskip 1em plus 0.5em minus 0.4em\relax Academic press, 1992, vol.~1.

\bibitem{guan2019wavelet}
J.~Guan, R.~Lai, and A.~Xiong, ``Wavelet deep neural network for stripe noise removal,'' \emph{IEEE Access}, vol.~7, pp. 44\,544--44\,554, 2019.

\bibitem{chang2019toward}
Y.~Chang, M.~Chen, L.~Yan, X.-L. Zhao, Y.~Li, and S.~Zhong, ``Toward universal stripe removal via wavelet-based deep convolutional neural network,'' \emph{IEEE Trans. Geosci. Remote Sens.}, vol.~58, no.~4, pp. 2880--2897, 2019.

\bibitem{zhang2021research}
S.~Zhang, X.~Sui, Z.~Yao, G.~Gu, and Q.~Chen, ``Research on nonuniformity correction based on deep learning,'' in \emph{AOPC 2021: Infrared Device and Infrared Technology}, vol. 12061.\hskip 1em plus 0.5em minus 0.4em\relax SPIE, 2021, pp. 97--102.

\bibitem{isola2017image}
P.~Isola, J.-Y. Zhu, T.~Zhou, and A.~A. Efros, ``Image-to-image translation with conditional adversarial networks,'' in \emph{Proc. IEEE/CVF Conf. Comput. Vis. Pattern Recognit. (CVPR)}, 2017, pp. 1125--1134.

\bibitem{zhu2017unpaired}
J.-Y. Zhu, T.~Park, P.~Isola, and A.~A. Efros, ``Unpaired image-to-image translation using cycle-consistent adversarial networks,'' in \emph{Proc. ICCV}, 2017, pp. 2223--2232.

\bibitem{lehtinen2018noise2noise}
J.~Lehtinen, J.~Munkberg, J.~Hasselgren, S.~Laine, T.~Karras, M.~Aittala, and T.~Aila, ``Noise2noise: Learning image restoration without clean data,'' \emph{arXiv preprint arXiv:1803.04189}, 2018.

\bibitem{mei2018unsupervised}
S.~Mei, H.~Yang, and Z.~Yin, ``An unsupervised-learning-based approach for automated defect inspection on textured surfaces,'' \emph{IEEE Trans. Instrum. Meas.}, vol.~67, no.~6, pp. 1266--1277, 2018.

\bibitem{song2020unsupervised}
J.~Song, J.-H. Jeong, D.-S. Park, H.-H. Kim, D.-C. Seo, and J.~C. Ye, ``Unsupervised denoising for satellite imagery using wavelet directional cyclegan,'' \emph{IEEE Trans. Geosci. Remote Sens.}, vol.~59, no.~8, pp. 6823--6839, 2020.

\bibitem{wei2021deraincyclegan}
Y.~Wei, Z.~Zhang, Y.~Wang, M.~Xu, Y.~Yang, S.~Yan, and M.~Wang, ``Deraincyclegan: Rain attentive cyclegan for single image deraining and rainmaking,'' \emph{IEEE Trans. Image Process.}, vol.~30, pp. 4788--4801, 2021.

\bibitem{lin2023deflickercyclegan}
X.~Lin, Y.~Li, J.~Zhu, and H.~Zeng, ``Deflickercyclegan: learning to detect and remove flickers in a single image,'' \emph{IEEE Trans. Image Process.}, vol.~32, pp. 709--720, 2023.

\bibitem{liu2023toward}
X.~Liu, T.~Zhang, and J.~Zhang, ``Toward visual quality enhancement of dehazing effect with improved cycle-gan,'' \emph{Neural Comput Appl.}, vol.~35, no.~7, pp. 5277--5290, 2023.

\bibitem{narayanan2005scene}
B.~Narayanan, R.~C. Hardie, and R.~A. Muse, ``Scene-based nonuniformity correction technique that exploits knowledge of the focal-plane array readout architecture,'' \emph{Appl. Opt.}, vol.~44, no.~17, pp. 3482--3491, 2005.

\bibitem{liu2015fixed}
Z.~Liu, J.~Xu, X.~Wang, K.~Nie, and W.~Jin, ``A fixed-pattern noise correction method based on gray value compensation for tdi cmos image sensor,'' \emph{Sensors}, vol.~15, no.~9, pp. 23\,496--23\,513, 2015.

\bibitem{zhao2016loss}
H.~Zhao, O.~Gallo, I.~Frosio, and J.~Kautz, ``Loss functions for image restoration with neural networks,'' \emph{IEEE Trans. Comput. Imaging}, vol.~3, no.~1, pp. 47--57, 2016.

\bibitem{park2019densely}
B.~Park, S.~Yu, and J.~Jeong, ``Densely connected hierarchical network for image denoising,'' in \emph{Proc. IEEE/CVF Conf. Comput. Vis. Pattern Recognit. (CVPR)}, 2019, pp. 0--0.

\bibitem{yin2023multiscale}
X.~Yin, G.~Tu, and Q.~Chen, ``Multiscale depth fusion with contextual hybrid enhancement network for image dehazing,'' \emph{IEEE Trans. Instrum. Meas.}, 2023.

\bibitem{li2022mafusion}
X.~Li, H.~Chen, Y.~Li, and Y.~Peng, ``M{AF}usion: Multiscale attention network for infrared and visible image fusion,'' \emph{IEEE Trans. Instrum. Meas.}, vol.~71, pp. 1--16, 2022.

\bibitem{misra2021rotate}
D.~Misra, T.~Nalamada, A.~U. Arasanipalai, and Q.~Hou, ``Rotate to attend: Convolutional triplet attention module,'' in \emph{Proc. IEEE/CVF Conf. Comput. Vis. Pattern Recognit. (CVPR)}, 2021, pp. 3139--3148.

\bibitem{johnson2016perceptual}
J.~Johnson, A.~Alahi, and L.~Fei-Fei, ``Perceptual losses for real-time style transfer and super-resolution,'' in \emph{Proc. Eur. Conf. Comput. Vis. (ECCV)}.\hskip 1em plus 0.5em minus 0.4em\relax Springer, 2016, pp. 694--711.

\bibitem{simonyan2014very}
K.~Simonyan and A.~Zisserman, ``Very deep convolutional networks for large-scale image recognition,'' \emph{arXiv preprint arXiv:1409.1556}, 2014.

\bibitem{deng2009imagenet}
J.~Deng, W.~Dong, R.~Socher, L.-J. Li, K.~Li, and L.~Fei-Fei, ``Imagenet: A large-scale hierarchical image database,'' in \emph{Proc. IEEE/CVF Conf. Comput. Vis. Pattern Recognit. (CVPR)}.\hskip 1em plus 0.5em minus 0.4em\relax Ieee, 2009, pp. 248--255.

\bibitem{lee2018diverse}
H.-Y. Lee, H.-Y. Tseng, J.-B. Huang, M.~Singh, and M.-H. Yang, ``Diverse image-to-image translation via disentangled representations,'' in \emph{Proc. Eur. Conf. Comput. Vis. (ECCV)}, 2018, pp. 35--51.

\bibitem{zhang2017beyond}
K.~Zhang, W.~Zuo, Y.~Chen, D.~Meng, and L.~Zhang, ``Beyond a gaussian denoiser: Residual learning of deep cnn for image denoising,'' \emph{IEEE Trans. Image Process.}, vol.~26, no.~7, pp. 3142--3155, 2017.

\bibitem{socarras2013adapting}
Y.~Socarr{\'a}s, S.~Ramos, D.~V{\'a}zquez, A.~M. L{\'o}pez, and T.~Gevers, ``Adapting pedestrian detection from synthetic to far infrared images,'' in \emph{ICCV Workshops}, vol.~3, 2013.

\bibitem{kuang2018robust}
X.~Kuang, X.~Sui, Y.~Liu, C.~Liu, Q.~Chen, and G.~Gu, ``Robust destriping method based on data-driven learning,'' \emph{Infr. Phys. Technol}, vol.~94, pp. 142--150, 2018.

\bibitem{huang2022winnet}
J.-J. Huang and P.~L. Dragotti, ``W{INN}et: Wavelet-inspired invertible network for image denoising,'' \emph{IEEE Trans. Image Process.}, vol.~31, pp. 4377--4392, 2022.

\bibitem{wang2004image}
Z.~Wang, A.~C. Bovik, H.~R. Sheikh, and E.~P. Simoncelli, ``Image quality assessment: from error visibility to structural similarity,'' \emph{IEEE Trans. Image Process.}, vol.~13, no.~4, pp. 600--612, 2004.

\end{thebibliography}

\begin{IEEEbiography}[{\includegraphics[width=1in,height=1.2in,clip,keepaspectratio]{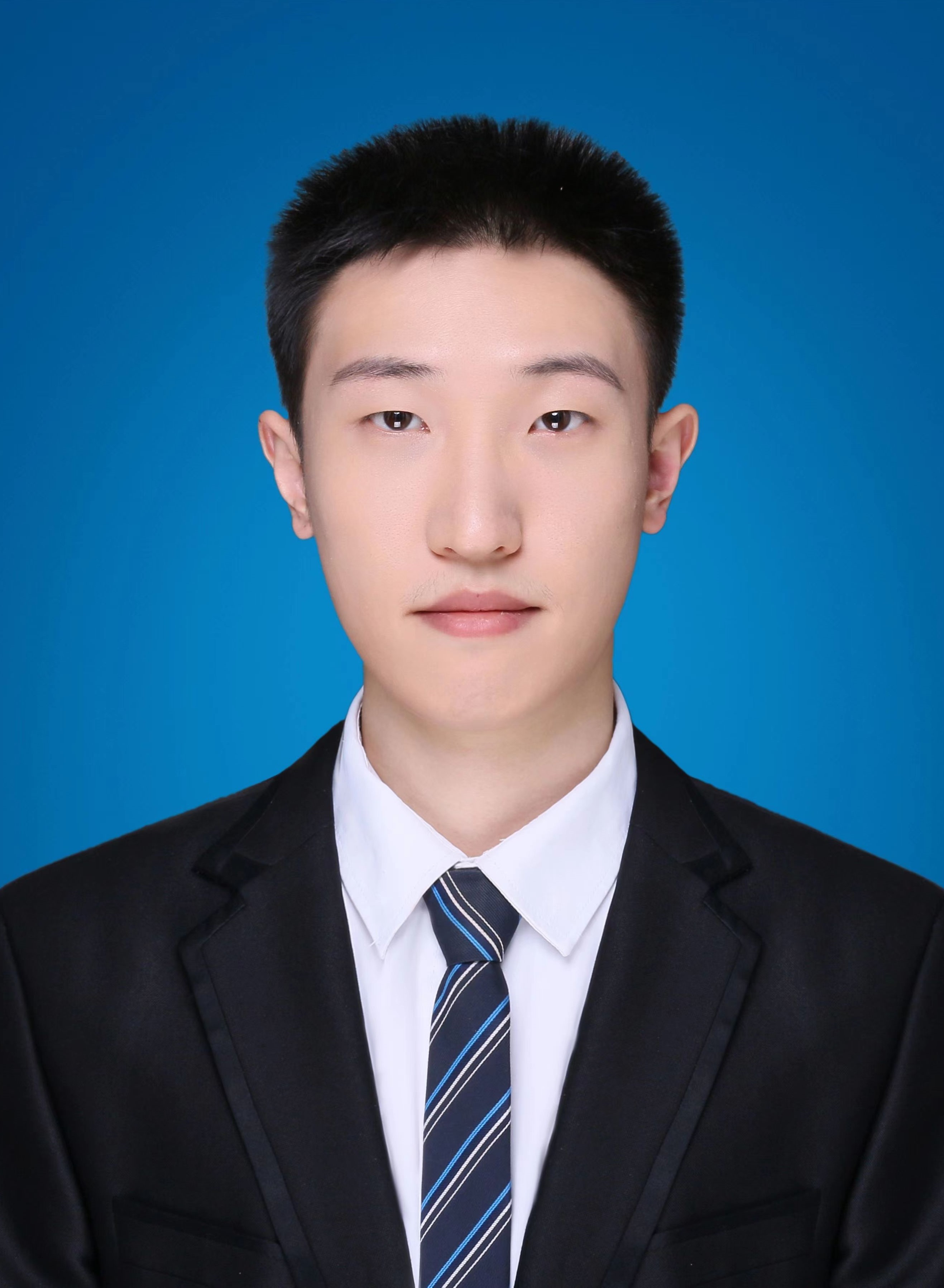}}]{Shiqi Yang}
received the B.S. degree in electronic science and technology from Xidian University, Shanxi, China, in 2022. He is currently pursuing the M.S. degree with Xidian University. His current research interests include deep learning, image processing, and image deraining.
\end{IEEEbiography}

\begin{IEEEbiography}[{\includegraphics[width=1in,height=1.2in,clip,keepaspectratio]{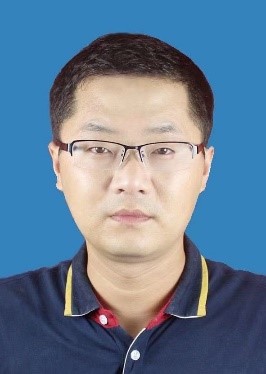}}]{Hanlin Qin}
received the B.S and Ph.D. degrees from Xidian University, Xi'an, China, in 2004 and 2010. He is currently a full professor with the School of Optoelectronic Engineering, Xidian University. He authored or co-authored more than 100 scientific articles. His research interests include electro-optical cognition, advanced intelligent computing, and autonomous collaboration.
\end{IEEEbiography}

\begin{IEEEbiography}[{\includegraphics[width=1in,height=1.13in,clip,keepaspectratio]{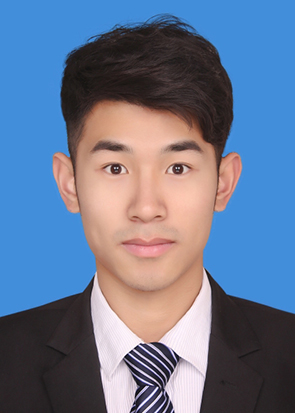}}]
{Shuai Yuan}
received the B.S. degree from Xi'an Technological University, Xi'an, China, in 2019. He is currently pursuing a Ph.D. degree at Xidian University, Xi’an, China. He currently studies at the University of Melbourne as a visiting student, working closely with Dr. Naveed Akhtar. His research interest includes infrared image understanding, remote sensing, and deep learning.
\end{IEEEbiography}

\begin{IEEEbiography}[{\includegraphics[width=1in,height=1.25in, clip,keepaspectratio]{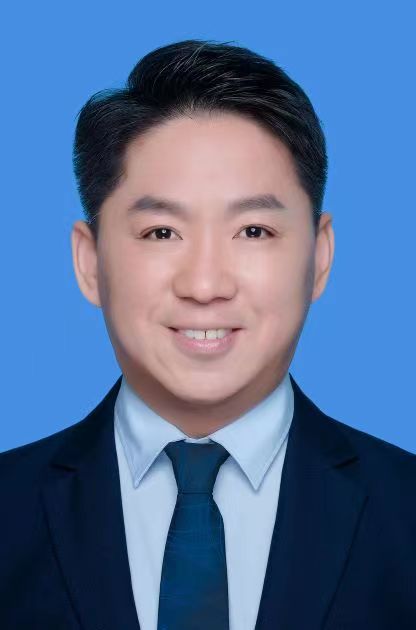}}]{Xiang Yan} 
received the B.S and Ph.D. degrees from Xidian University, Xi'an, China, in 2012 and 2018. He was a visiting Ph.D. Student with the School of Computer Science and Software Engineering, Australia, from 2016 to 2018, working closely with Prof. Ajmal Mian. He is currently an associate professor at Xidian University, Xi'an, China. His current research interests include image processing, computer vision and deep learning.
\end{IEEEbiography}

\vspace{-125mm}
\begin{IEEEbiography}[{\includegraphics[width=1in,height=1.25in, clip,keepaspectratio]{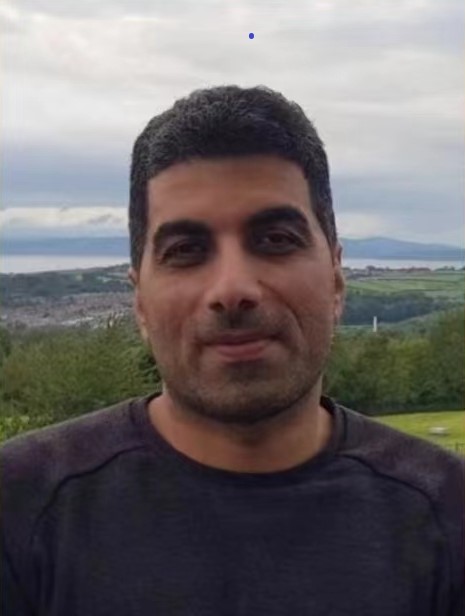}}]{HosseinRahmani} 
received the B.Sc. degree in computer software engineering from the Isfahan University of Technology, Isfahan, Iran, in 2004, the M.Sc. degree in software engineering from Shahid Beheshti University, Tehran, Iran, in 2010, and the Ph.D. degree from The University of Western Australia, Perth, WA, Australia, in 2016. He is an Associate Professor (Senior Lecturer) with the School of Computing and Communications at Lancaster University in the UK. Before that, he was a Research Fellow with the School of Computer Science and Software Engineering, The University of Western Australia. His research interests include computer vision, action recognition, pose estimation, and deep learning.
\end{IEEEbiography}

\end{document}